\documentclass[superscriptaddress,pdftex,twocolumn,epjc3]{svjour3}

\usepackage[switch]{lineno}
\usepackage[slovak,english]{babel}
\usepackage{subfig}
\usepackage[utf8x]{inputenc}
\usepackage[T1]{fontenc}
\usepackage{mathptmx}
\usepackage{amsmath}
\usepackage{amssymb}
\usepackage[tight]{units}
\usepackage{color,graphicx}
\usepackage{threeparttable}
\usepackage{booktabs}
\usepackage{upgreek}
\usepackage{multirow}
\usepackage[%
separate-uncertainty = true,
table-align-text-post = false
]{siunitx}
\usepackage{textcomp}
\usepackage{mathtools}
\usepackage[numbers,sort&compress]{natbib}
\usepackage[colorlinks,citecolor=blue,urlcolor=blue,linkcolor=blue]{hyperref}
\graphicspath{{images/}}
\DeclareGraphicsExtensions{.pdf,.png,.jpg}
\usepackage{appendix}

\usepackage{color, colortbl}
\definecolor{Grey}{rgb}{0.5,0.5,0.5}
\definecolor{Purple}{rgb}{1,0.2,1}
\definecolor{Blue}{rgb}{0,0.5,1}
\definecolor{Green}{rgb}{0,1,0}
\definecolor{Orange}{rgb}{1,0.6,0.2}
\definecolor{Red}{rgb}{1,0.2,0.2}
\definecolor{LightCyan}{rgb}{0.88,1,1}

\definecolor{LightGrey}{RGB}{212,214,214}
\definecolor{LightPurple}{RGB}{250,205,245}
\definecolor{LightBlue}{RGB}{205,218,250}
\definecolor{LightGreen}{RGB}{205,250,216}
\definecolor{LightOrange}{RGB}{250,236,205}
\definecolor{LightRed}{RGB}{250,205,205}
\definecolor{LightLightCyan}{RGB}{205,250,247}

\begin{document}
\author{
  A.H.~Abdelhameed\thanksref{addrMPI}\and
  G.~Angloher\thanksref{addrMPI}\and
  P.~Bauer\thanksref{addrMPI}\and
  A.~Bento\thanksref{addrMPI,addrCoimbra}\and 
  E.~Bertoldo\thanksref{addrMPI}\and
  R.~Breier\thanksref{addrBratislava}\and
  C.~Bucci\thanksref{addrLNGS}\and 
  L.~Canonica\thanksref{addrMPI}\and 
  A.~D'Addabbo\thanksref{addrLNGS,addrGSSI}\and
  S.~Di~Lorenzo\thanksref{addrLNGS,addrGSSI}\and
  A.~Erb\thanksref{addrTUM,addrWMI}\and
  F.~v.~Feilitzsch\thanksref{addrTUM}\and 
  N.~Ferreiro~Iachellini\thanksref{addrMPI}\and
  S.~Fichtinger\thanksref{addrHEPHY}\and
  A.~Fuss\thanksref{addrHEPHY,addrAI}\and
  P.~Gorla\thanksref{addrLNGS}\and 
  D.~Hauff\thanksref{addrMPI}\and 
  M.~Ješkovsk\'y\thanksref{addrBratislava}\and
  J.~Jochum\thanksref{addrTUE}\and
  J.~Kaizer\thanksref{addrBratislava}\and
  A.~Kinast\thanksref{addrTUM}\and
  H.~Kluck\thanksref{e1,addrHEPHY,addrAI}\and
  H.~Kraus\thanksref{addrOxford}\and 
  A.~Langenk\"amper\thanksref{addrTUM}\and 
  M.~Mancuso\thanksref{addrMPI}\and
  V.~Mokina\thanksref{e2,addrHEPHY}\and
  E.~Mondrag\'on\thanksref{addrTUM}\and
  M.~Olmi\thanksref{addrLNGS,addrGSSI}\and
  T.~Ortmann\thanksref{addrTUM}\and
  C.~Pagliarone\thanksref{addrLNGS,addrCASS}\and
  V.~Palušov\'a\thanksref{addrBratislava}\and
  L.~Pattavina\thanksref{addrTUM,addrGSSI}\and
  F.~Petricca\thanksref{addrMPI}\and 
  W.~Potzel\thanksref{addrTUM}\and 
  P.~Povinec\thanksref{addrBratislava}\and
  F.~Pr\"obst\thanksref{addrMPI}\and     
  F.~Reindl\thanksref{addrHEPHY,addrAI} \and
  J.~Rothe\thanksref{addrMPI}\and 
  K.~Sch\"affner\thanksref{addrMPI}\and 
  J.~Schieck\thanksref{addrHEPHY,addrAI}\and 
  V.~Schipperges\thanksref{addrTUE}\and
  D.~Schmiedmayer\thanksref{addrHEPHY,addrAI}\and
  S.~Sch\"onert\thanksref{addrTUM}\and 
  C.~Schwertner\thanksref{addrHEPHY,addrAI}\and
  M.~Stahlberg\thanksref{addrHEPHY,addrAI}\and 
  L.~Stodolsky\thanksref{addrMPI}\and 
  C.~Strandhagen\thanksref{addrTUE}\and
  R.~Strauss\thanksref{addrTUM}\and
  C.~T\"urko\u{g}lu\thanksref{e3,addrHEPHY,addrAI,addrSussex}\and
  I.~Usherov\thanksref{addrTUE}\and 
  M.~Willers\thanksref{addrTUM}\and 
  V.~Zema\thanksref{addrLNGS,addrGSSI,addrGOTE}\and
  J.~Zeman\thanksref{addrBratislava}\\
(CRESST Collaboration)
}

\institute
{Max-Planck-Institut f\"ur Physik, D-80805 M\"unchen, Germany \label{addrMPI} \and
Comenius University, Faculty of Mathematics, Physics and Informatics, 84248 Bratislava, Slovakia \label{addrBratislava} \and
INFN, Laboratori Nazionali del Gran Sasso, I-67100 Assergi, Italy \label{addrLNGS} \and
Physik-Department and Excellence Cluster Universe, Technische Universit\"at M\"unchen, D-85747 Garching, Germany \label{addrTUM} \and
Institut f\"ur Hochenergiephysik der \"Osterreichischen Akademie der Wissenschaften, A-1050 Wien, Austria\label{addrHEPHY} \and
Atominstitut, Technische Universit\"at Wien, A-1020 Wien, Austria \label{addrAI} \and
Eberhard-Karls-Universit\"at T\"ubingen, D-72076 T\"ubingen, Germany \label{addrTUE} \and
Department of Physics, University of Oxford, Oxford OX1 3RH, United Kingdom \label{addrOxford} \and
also at: Departamento de Fisica, Universidade de Coimbra, P3004 516 Coimbra, Portugal \label{addrCoimbra} \and
also at: GSSI-Gran Sasso Science Institute, I-67100 L'Aquila, Italy \label{addrGSSI} \and
also at: Walther-Mei\ss ner-Institut f\"ur Tieftemperaturforschung, D-85748 Garching, Germany \label{addrWMI} \and
also at: Dipartimento di Ingegneria Civile e Meccanica, Universitá degli Studi di Cassino e del Lazio Meridionale, I-03043 Cassino, Italy\label{addrCASS} \and
also at: Chalmers University of Technology, Department of Physics, SE-412 96 Göteborg, Sweden \label{addrGOTE} \and
present address: University of Sussex, School of Mathematical and Physical Sciences, BN1 9QH Brighton, United Kingdom \label{addrSussex}
}

\thankstext{e1}{Corresponding author: holger.kluck@oeaw.ac.at}
\thankstext{e2}{Corresponding author: valentyna.mokina@oeaw.ac.at}
\thankstext{e3}{Corresponding author: c.turkoglu@sussex.ac.uk}

\title{Geant4-based electromagnetic background model for the CRESST dark matter experiment}
\maketitle

\begin{abstract}
The CRESST (Cryogenic Rare Event Search with Superconducting Thermometers) dark matter search experiment aims for the detection of dark matter particles via elastic scattering off nuclei in $\mathrm{CaWO_4}$ crystals. To understand the CRESST electromagnetic background due to the bulk contamination in the employed materials, a model based on Monte Carlo simulations was developed using the Geant4 simulation toolkit. The results of the simulation are applied to the TUM40 detector module of CRESST-II phase 2. We are able to explain up to
$(68 \pm 16)\,\mathrm{\%}$ of the electromagnetic background in the energy range between $1\,\mathrm{keV}$ and $40\,\mathrm{keV}$.
\end{abstract}

\section{Introduction}
Understanding the nature of dark matter (DM) is among the most pressing problems of modern physics. The possible parameter space in terms of mass and interaction cross section of ordinary matter with DM is large, requiring different experimental approaches to explore it in a comprehensive way. The cryogenic experiment CRESST is one of these experiments, being very well suited for the search for DM particles in the sub-$\mathrm{GeV/c^2}$ mass region, down to a few hundred $\mathrm{MeV/c^2}$~\cite{Abdelhameed:2019hmk}.
A major milestone for the CRESST-II experiment in increasing the sensitivity to sub-$\mathrm{GeV/c^2}$ dark matter was the data taking campaign of phase 2: with a detection threshold of $(307.3 \pm 3.6)\,\mathrm{eV}$, the single detector module Lise achieved sensitivity down to $500\,\mathrm{MeV/c^2}$ \cite{Angloher:2016}. This demonstrates the importance of a sub-keV detection threshold for nuclear recoils. The first time, it was achieved earlier in the same run with the detector module TUM40, for which a threshold of $(603\pm2)\,\mathrm{eV}$ was obtained \cite{Angloher:2014myn}. 

The average background rate previously determined in the crystal used for the TUM40 detector module in the region 1 - 40 keV of about $3.51\, \mathrm{kg^{-1} keV^{-1} d^{-1}}$ \cite{Strauss:2014aqw} is significantly lower compared to crystals of the same material with different origin.
The sensitivity for low DM masses was limited by the residual background and detection threshold, hence, further improvements are only possible by identifying and reducing components of this background or lowering the detection threshold.

In this paper we present an analysis of the electromagnetic background sources and their composition \cite{dissertation_cenk} for the data taken during CRESST-II phase 2 with the TUM40 detector module \cite{Angloher:2014myn,Strauss:2014aqw}. We use the Geant4 toolkit for Monte Carlo simulations \cite{Agostinelli2003, Allison2006, Allison2016} as the main tool to calculate the differential energy distribution from the decay of various radioactive isotopes. 
The normalisation of the energy distribution is obtained from the activity of $\alpha$-decays in the corresponding natural decay chain in secular equilibrium or from the fit to $\gamma$-lines. \par
In the following, we present the CRESST experiment in Section \ref{cresstexperiment}. Section \ref{montecarlosim} focuses on the simulation method and Section \ref{sec:4} discusses the determination of the different background contributions. The results are presented in Section \ref{sec:results}. Section \ref{sec:summary} gives a summary and provides a short outlook. 

\section{The CRESST experiment} \label{cresstexperiment}
\subsection{The working principle}
CRESST experiment is located in the hall A of the Laboratori Nazionali del Gran Sasso (LNGS) below the Gran Sasso mountains in central Italy. CRESST-II used cryogenic detectors with scintillating $\mathrm{CaWO_4}$ crystals as the target material. Each interaction in the target crystal produces phonons and scintillation photons, which are absorbed by a nearby silicon-on-sapphire (SOS) disk. Around $90\,\mathrm{\%}$ of the total energy deposited in the target crystal are phonons and nearly independent of the interacting particle. On the other hand, the scintillation process is strongly particle-dependent. Its efficiency reaches at most $7\,\mathrm{\%}$ for $\mathrm{\beta/\gamma}$-events \cite{Kiefer:2015sha}; for nuclear recoils the scintillation efficiency is reduced by roughly one order of magnitude called quenching. Both signals are read out by separate thermometers which are realised as transition edge sensors (TESs) operated at a temperature of $\mathcal{O}(15\,\mathrm{mK})$ \cite{Pobell}. We refer to the $\mathrm{CaWO_4}$ target crystal as “\textit{phonon detector}” and the SOS disk as “\textit{light detector}” later on. The ratio between deposited energies in these two channels, defines as the light yield (\textit{LY}), allows to determine the type of interaction.
The $122\,\mathrm{keV}$ $\gamma$-line coming from a $\mathrm{{}^{57}Co}$ calibration source is used to normalise the LY of $ \beta$/$\gamma$-band to 1 at $122\,\mathrm{keV}$. \par
\begin{figure}[tbp]
\includegraphics[width=\linewidth]{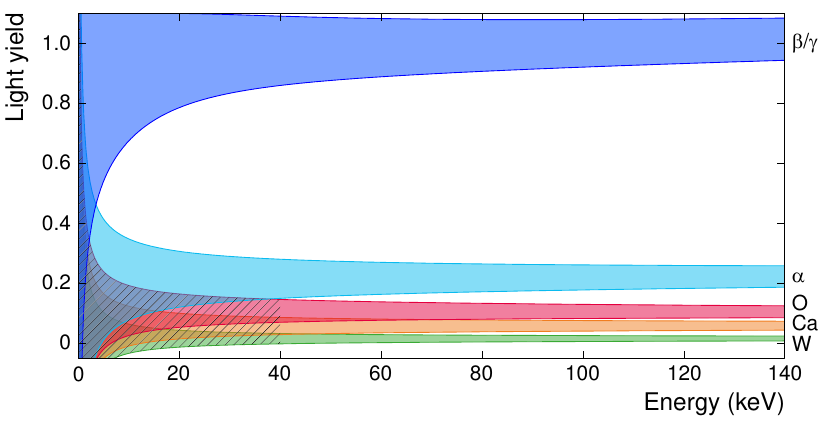}
\caption{An example LY plot. Different bands show interactions of different event types: $ \beta$/$\gamma$-interactions (\textit{blue}), $\alpha$ interactions (\textit{cyan}), oxygen (\textit{red}), calcium (\textit{orange}) and tungsten (\textit{green}) recoils. The \textit{hatched area} shows the region of interest for a possible signal of DM particles.}
\label{fig:light_yield}
\end{figure}

Fig.~\ref{fig:light_yield} shows the expected LY distribution as a function of the deposited energy. As can be seen, a number of bands emerge, separated by their respective LYs. Different bands indicate the different particle interactions inside the target crystal. Starting from high LY values to lower ones, the resulting event categories are: $\beta$/$\gamma$-events caused by scattering off electrons, $\alpha$-events and finally events of recoiling $\mathrm{O}$-, $\mathrm{Ca}$- and $\mathrm{W}$-nuclei. The latter three collectively constitute the nuclear recoil bands. A recoil event may be caused by background, e.g.\ neutrons, or by a potential DM signal.
The energy range of interest (ROI) for a DM search in CRESST-II expands from the detection threshold to $40\,\mathrm{keV}$. Above this energy, no significant DM signal is expected for $\mathrm{CaWO_4}$ \cite{Angloher:2014myn}.\par
Most of the radioactive background falls into the $ \beta$/$\gamma$-band which is well separated from the nuclear recoil bands at high energies. However, at lower energies, events from the $ \beta$/$\gamma$-band are leaking into the region of interest. This complicates the DM analysis: since the expected rate of DM particle interactions is low compared to the background rate, a potential signal may be covered by the leakage, or an unknown background can be mistaken for a potential DM signal. Especially at low energies where the ROI for CRESST experiment is, a detailed understanding of the background is crucial.

\subsection{The detector setup} \label{setup}
\begin{figure}[ht!]
\includegraphics[width=\linewidth]{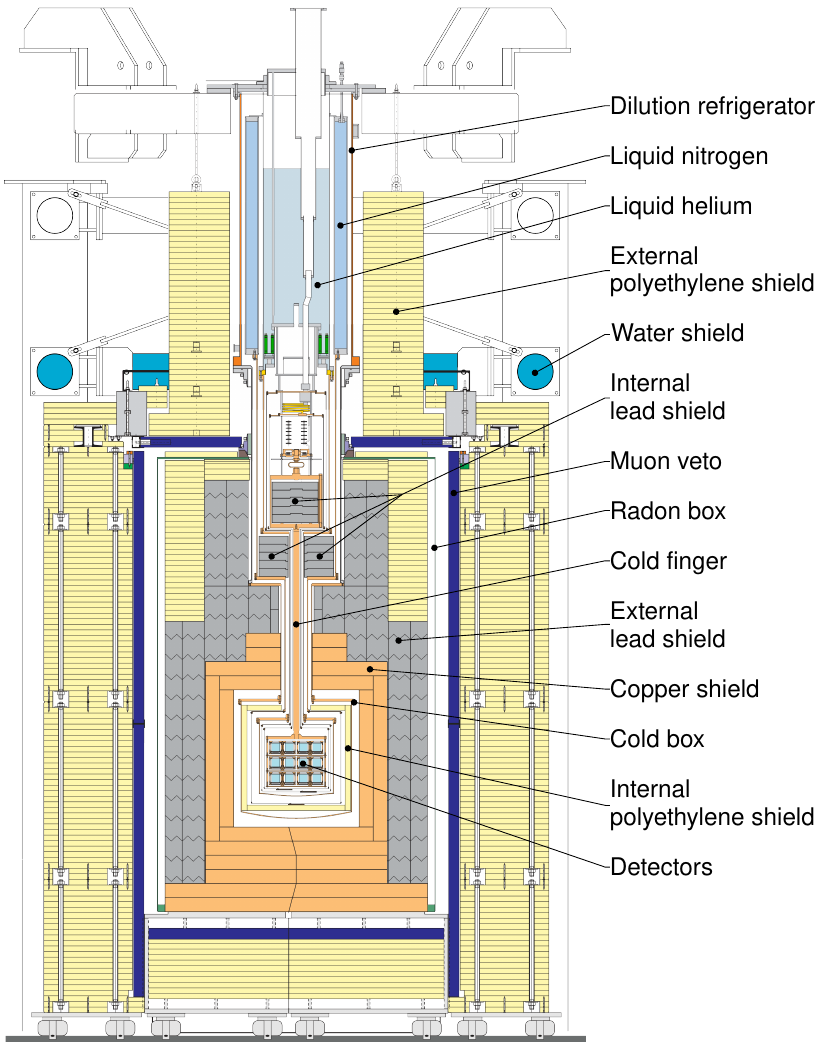}
\caption{Technical drawing of the CRESST experimental setup. The cryostat holding the liquid helium and nitrogen can be seen at the top. Below the cryostat, the carousel hosts the detector modules. The shielding consists of PE, lead, copper, an active muon veto and the air-tight radon box used to prevent radon contamination from air.}
\label{fig:set_up}
\end{figure}

Most of the materials close to the detectors are selected for radiopurity to keep backgrounds low. The cold finger design of CRESST helps to avoid any line of sight between the dilution refrigerator and the detectors ~\cite{Angloher:2002} (see Fig.~\ref{fig:set_up}). The design of the CRESST setup features layers of polyethylene (PE), lead and copper shielding against ambient background from natural radioactivity of the laboratory, as well as background caused by the shielding and cryostat itself (see Section \ref{bgsources} for details). \par
The active muon veto tags when a charged particle penetrates through it. A gas-tight container, the so-called radon box, which is constantly flushed with $\mathrm{N_2}$ gas, prevents the accumulation of gaseous radon close to the detectors. The radon box contains a lead shielding, inside this a copper shielding is the direct enclosing of the cold box. The detector modules are placed in the so-called carousel at the core of the coldbox.\par
\begin{figure}[htbp]
\subfloat[]{
    \label{fig:tum_photo}
    \includegraphics[width=\linewidth]{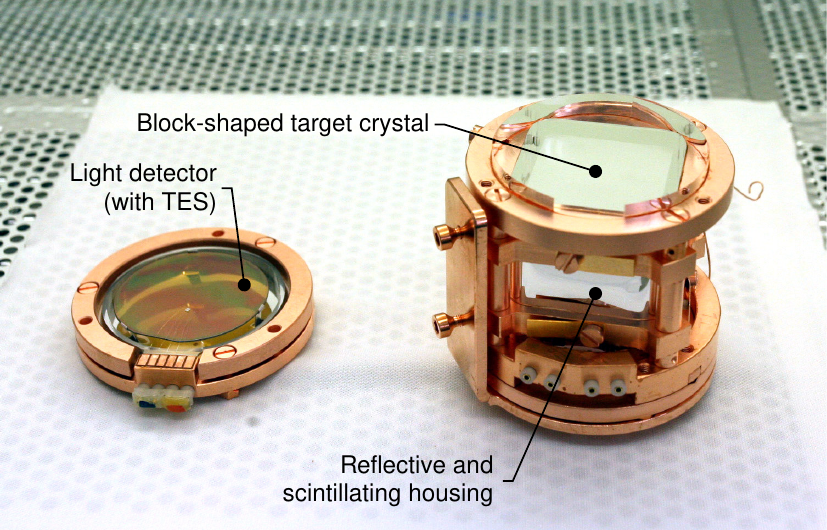}
}\\
\subfloat[]{
    \label{fig:tum_sketch}
    \includegraphics[width=\linewidth]{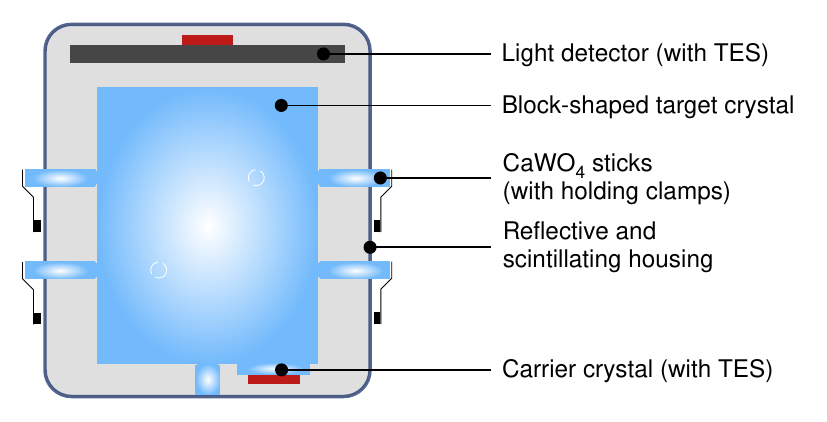}
}\\
\caption{TUM40 detector module as operated in CRESST-II Phase 2: \protect\subref{fig:tum_photo} photograph of opened module, the block-shaped target crystal is visible on the right side within the scintillating foil, and \protect\subref{fig:tum_sketch} a schematic view.}
\label{fig:tum40}
\end{figure}

A photograph and schematic view of TUM40 detector module \cite{Strauss:2014hog} are shown in Fig.~\ref{fig:tum_photo} and \ref{fig:tum_sketch}, respectively. The target crystal is held by sticks made of the same material - $\mathrm{CaWO_4}$ \cite{Strauss:2014hog} and features a dedicated carrier crystal onto which the TES is evaporated. The phonon and light detectors are enclosed by a reflective scintillating polymeric foil named VM2002\footnote{Product nowadays continued under the name Vikuiti, produced by the company 3M, \url{https://www.3m.com}}, to increase the light collection efficiency and actively reject surface backgrounds. The holder of the detector module is made from radiopure copper. A more detailed description of the TUM40 geometry is given in Section \ref{geant4}, where the implementation into the simulation code is discussed.

\subsection{Potential background sources} \label{bgsources}
In this Section, the main background sources and the methods to reduce and identify them will be discussed. For the TUM40 detector module, previous studies \cite{Angloher:2014myn, Strauss:2014aqw} found no indication for a significant neutron component to the background. Based on CRESST-II phase 1 we expect $<10^{-3}\,\mathrm{kg^{-1}\,keV^{-1}\,d^{-1}}$ of neutron background \cite{Angloher:2011uu}. Hence, we focus on the electromagnetic components in this work.

\subsubsection{Muons}
High energetic atmospheric muons can easily penetrate through meters-thick layers of shielding. For this reason, the CRESST experiment is located in the LNGS underground laboratory, which has an overburden of at least $1400\,\mathrm{m}$ of rock in each direction. This provides an average shielding of $3600\,\mathrm{m}$ w.\ e.\ \cite{Ambrosio:1995cx,Selvi}. The remaining muon rate is $\mathcal{O}(1\,\mathrm{h^{−1}m^{−2}})$ \cite{Ambrosio:1995cx}, i.e. it is reduced by about six orders of magnitude compared to that at sea level.\par
The remaining muons can induce background events in detectors, either by direct interaction with the target crystal, or by producing secondary particles in their interactions with rock or the material of the experimental setup, e.g. bremsstrahlung or highly energetic neutrons by muon-induced spallation. To be able to reject muon-related events, CRESST is equipped with a muon veto (see Fig.~\ref{fig:set_up}). The veto consists of 20 plastic scintillator panels, read out by photomultiplier tubes and covering $98.7\,\%$ of the solid angle around the detectors. The reason why the coverage is not $100\,\%$ is that an opening is needed on the top to leave space for the cryostat. If the muon veto triggered, all events recorded by cryogenic detectors, within a time window $\pm 2\,\mathrm{ms}$, are rejected in the offline analysis.

\subsubsection{Cosmogenic activation}\label{sec:cosmicActivation}
Long-lived radioactive impurities in the detector material might be activated due to its exposure to cosmic rays before being brought underground \cite{Cebrian:2017oft}. For $\mathrm{CaWO_4}$, lines caused by the decays of cosmogenically activated nuclides below $80\,\mathrm{keV}$ are reported in \cite{Lang:2009wb,Strauss:2014aqw,Angloher:2016,Abdelhameed:2019hmk}. \par
Four lines can be detected due to cosmogenic activation of $^{182}\mathrm{W}$ via proton capture. With a half-life of $T_{1/2}=1.82\,\mathrm{yr}$ \cite{iaea} the resulting $\mathrm{^{179}Ta}$ then decays in the target material by electron capture (EC), leading to the occurrence of the X-ray line as follows:
\begin{equation} \label{182W}
\begin{alignedat}{2}
\mathrm{^{182}W} + \mathrm{p} \longrightarrow & \mathrm{^{179}Ta} + \mathrm{\alpha}  \\ 
                  {}                          & \hookrightarrow \mathrm{^{179}Ta} + \mathrm{e^-} & \longrightarrow & \mathrm{^{179}Hf^{*}} + \mathrm{\upsilon_{e}}  \\
                  {}                          &             {}                 &        {}      & \hookrightarrow \mathrm{^{179}Hf} + \text{X-ray}.
\end{alignedat}
\end{equation}
Since the decay happens in the target crystal the entire atomic de-excitation is measured. The signature of the EC therefore is a line at the binding energy of the respective captured shell electron of $^{179}\mathrm{Hf}$. Peaks originating from this decay can be seen at energies of $2.60\,\mathrm{keV}$ ($\mathrm{M_1}$-shell), $10.74\,\mathrm{keV}$ ($\mathrm{L_2}$-shell\footnote{We note that the $L_1/L_2$ ratio does not match the expectation \cite{Firestone1996}, hence there may be a contribution of unknown background or there are opportunities for future enhancements.}), $11.27\,\mathrm{keV}$ ($\mathrm{L_1}$-shell) and $65.35\,\mathrm{keV}$ ($\mathrm{K}$-shell) \cite{Strauss:2014aqw}. \par
Another cosmogenically activated nuclide is $^{183}\mathrm{W}$ thro\-ugh $\mathrm{{}^{183}W(p,{}^{3}H)}$, resulting in $\mathrm{{}^{181}W}$. This then decays with $T_{1/2}=121.2\,\mathrm{d}$ \cite{iaea} into an excited state of $\mathrm{{}^{181}Ta}$ through EC and it de-excites with the emission of a $6.2\,\mathrm{keV}$ $\gamma$-ray:
\begin{equation} \label{183W}
\begin{alignedat}{2}
\mathrm{^{183}W} + \mathrm{p} \longrightarrow & \mathrm{^{181}W} + \mathrm{^{3}H}  & {}              & {}\\
                 {}                           & \hookrightarrow \mathrm{^{181}W} + \mathrm{e^-}               & \longrightarrow & \mathrm{^{181}Ta^{*}} + \mathrm{\upsilon_{e}}  \\
                 {}                           &            {}                                               &      {}         & \mathmakebox[0pt][l]{\hookrightarrow \mathrm{^{181}Ta} + \gamma + \text{X-ray}.}
\end{alignedat}
\end{equation}
Due to the slow detector response, the $\gamma$-ray and the atomic de-excitation cannot be resolved. The most prominent peak for this decay is thus observed at $73.6\,\mathrm{keV}$ which is the sum of the binding energy of the K-shell electron of tantalum with an energy $67.4\,\mathrm{keV}$ and $6.2\,\mathrm{keV}$ from the subsequent nuclear de-excitation \cite{Lang:2009wb}.

A nuclide which can be cosmogenically produced in a wide range of target materials is $\mathrm{{}^{3}H}$ \cite{Zhang:2016rlz}. It is hard to be observed experimentally for $\mathrm{CaWO_4}$ in CRESST, since the $\beta$-spectrum of $\mathrm{{}^{3}H}$ is buried below the other low-energy background components. However, Eq.~\ref{183W} indicates that the observed $\mathrm{{}^{181}W}$ should be accompanied by a $\mathrm{{}^{3}H}$ component. Other reaction paths to create $^{3}\mathrm{H}$ in $\mathrm{CaWO_4}$ may also be possible.

\subsubsection{Ambient $\gamma$-radiation} \label{ambient_gamma}
Another type of background are ambient $\gamma$-rays, originating from the rock and concrete of the LNGS. The sources of this radiation are $\gamma$-decays from the natural radioactive decay chains of $\mathrm{{}^{232}Th}$, $\mathrm{{}^{235}U}$ and $\mathrm{{}^{238}U}$, but also individual radioactive nuclides like $\mathrm{{}^{40}K}$. The integrated $\gamma$-ray flux in hall A of LNGS for energies below $3\,\mathrm{MeV}$ was measured to be $(0.28 \pm 0.02)\,\mathrm{cm^{−2} \, s^{−1}}$ \cite{Haffke:2011fp}.\par
To shield the experiment against this radiation, CRESST detectors are surrounded by an external lead shielding with a mass of $24\,\mathrm{tonnes}$ and a thickness of $20\,\mathrm{cm}$ (see Fig.~\ref{fig:set_up}). With its high density and atomic number, lead is ideally suited to absorb $\gamma$-rays. However, it is not a very radiopure material, in particular its naturally occurring radioactive isotope $^{210}\mathrm{Pb}$ contributes to the background observed in the experiment. Therefor, CRESST lead was produced by the Swedish company Boliden\footnote{\url{https://www.boliden.com}} with a low $\mathrm{{}^{210}Pb}$ contamination of $35\,\mathrm{Bq \, kg^{-1}}$ \cite{lang_thesis}. In addition, an internal lead shielding is placed inside the cryostat directly below the mixing chamber and consisting of $330\,\mathrm{kg}$ of lead produced by the company Plombum\footnote{Plombum Firma-Laboratorium, ul. Rozrywka 20, 31-419 Kraków, Poland} in Poland with an even lower $\mathrm{{}^{210}Pb}$ contamination of only $3.6\,\mathrm{Bq \, kg^{-1}}$ \cite{lang_thesis}. 
To reduce any remaining $\gamma$-background coming from the lead, low-background copper with a mass of $10\,\mathrm{tonnes}$ and a thickness of $14\,\mathrm{cm}$ is placed between the cold box, which contains the detectors, and the lead shielding (see Fig.~\ref{fig:set_up}). All the copper used in the experiment is so-called NOSV copper, produced by the Norddeutsche Affinerie AG\footnote{Aurubis AG since 12 February 2008, \url{https://www.aurubis.com/en}}. Copper is still a good gamma attenuator and in contrast to lead it can be produced with a very low level ($<1\, \mathrm{mBq \, kg^{-1}}$) of internal radioactivity as it has no naturally occurring radioactive isotopes. This makes copper a more favourable shielding material closer to the detectors.\par
In addition, copper is also used for structures inside the cold box, such as holders for the detectors of the CRESST experiment due to its good thermal conductivity. The ambient $\gamma$-flux is reduced to negligible amounts by the described shielding structure. However, even if copper can be produced with low contamination levels, it remains a non-negligible background source due to its proximity to the detectors and large overall mass \cite{Strauss:2014aqw}.

\subsubsection{Internal contamination of $\mathrm{CaWO_4}$}
All $\mathrm{CaWO_4}$ crystals in CRESST-II Phase 1 and most in Phase 2 were commercial crystals produced by the Scientific Research Company “Carat”\footnote{\url{http://en.carat.electron.ua}} and the Prokhorov General Physics Institute of the Russian Academy of Sciences\footnote{\url{http://www.gpi.ru/eng}}. The contamination of these crystals by $\alpha$-emitters from the natural decay chains in the energy range between $1.5\,\mathrm{MeV}$ and $\sim 7\,\mathrm{MeV}$ was measured to be between $(3.05 \pm 0.02)\,\mathrm{mBq \, kg^{−1}}$ and $(107.13 \pm 0.14)\,\mathrm{mBq \, kg^{−1}}$ \cite{Munster:2014mga}. Even though the energies of $\alpha$-decays are above the ROI for the DM searches, the subsequent decay chains contain also $\beta$/$\gamma$-decays that deposit energy in the ROI.\par
To reduce the intrinsic background, it was decided to produce crystals within the collaboration \cite{Munster:2014mga} using a Czochralski furnace dedicated to grow only $\mathrm{CaWO_4}$ crystals \cite{C2CE26554K}. This approach allows to control all production stages to prevent contamination during crystal growth and after-growth treatments. One of these self-grown crystals was used for the detector module TUM40 (see Fig.~\ref{fig:tum40}). In comparison with commercial crystals, a factor of 2 to 10 decrease in the $ \beta$/$\gamma$-background in the energy range of $1-40\,\mathrm{keV}$ for TUM40 was achieved. The $\beta$- and $\gamma$-emissions, e.g.\ caused by the decay of $\mathrm{{}^{210}Pb}$, are also reduced since they originate from the same natural decay chains as the measured $\alpha$-decays.\par
Even though the amount of background reduction with TUM40 is significant, there still remains a non-negligible amount of $ \beta$/$\gamma$-background at low energies. Hence, it is of great importance to understand the contribution of each background source using the experimental data at hand. The knowledge of background components also helps to guide the efforts of background reduction.

\subsection{The reference data sets} \label{reference_data}
 As experimental reference for the simulation model described in Section~\ref{montecarlosim}, we are using the complete data\footnote{We note that for the DM analysis in \cite{Angloher:2014myn}, only $29.35\,\mathrm{kg \, d}$ of these data were used; the uncertainty on the exposures is negligible. For this reduced data set \cite{Strauss:2014aqw} states an average gross rate in the region $1-40\,\mathrm{keV}$ of about $3.51\, \mathrm{kg^{-1} keV^{-1} d^{-1}}$.} recorded with the TUM40 module during CRESST-II phase 2 with a gross exposure of $129.9\,\mathrm{kg \, d}$.  This data set contains an average gross rate in the region $1-40\,\mathrm{keV}$ of about $3.1\, \mathrm{kg^{-1} \, keV^{-1} \, d^{-1}}$. If the deposited energy is high enough, the detector gets into normal conducting state and the recorded pulses are saturated. Hence, the standard energy calibration, which is based on the determination of pulse amplitude, is less precise for high energies \cite{florian}. Therefore, we are using three separated data sets: \textit{low}, \textit{medium} and \textit{high}. The main properties of these reference data sets are listed in Table~\ref{tab:reference data}.\par
\begin{figure}[htbp]
\includegraphics[width=\linewidth]{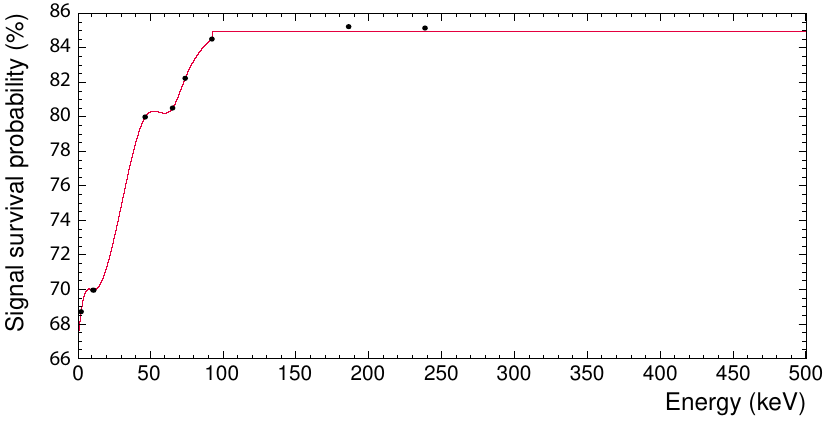}
\caption{Signal survival probability for the low-energy reference data. Up to the energy of $92.36\,\mathrm{keV}$, the signal survival probability is energy-dependent; above this energy, it is set to the constant value of 85.0\,\%.}
\label{fig:signal_survival}
\end{figure}

 To get the energy spectra for each data set different types of cuts are applied: a \textit{rate and stability cut}, a \textit{coincidence cut}, a \textit{carrier cut}, \textit{quality cuts} and \textit{RMS cuts} (see \cite{dissertation_cenk} for details). A detailed description of the cut types and their purpose are given in \cite{florian}. Successive applying the cuts used in the analysis gives the energy-dependent signal survival probability $\mathrm{\varepsilon}$. The latter is evaluated from simulated events as described in \cite{florian} and shown in Fig.~\ref{fig:signal_survival}. For the medium- and high-energy data sets the survival probabilities are energy-independent and listed in Table~\ref{tab:reference data}.\par
 \begin{table}
     \centering
     \caption{Properties of reference data sets: energy range of measured data before $E$ and after rescaling $E'$, signal survival probability $\mathrm{\varepsilon}$, and net number of background events $N$.}
     \label{tab:reference data}
    \begin{threeparttable}
    \begin{tabular*}{\columnwidth}{@{\extracolsep{\fill}}lccS[table-format=2.1(0)]S[table-format=4.0(0)]}
     \hline
        Data Sets & $E$ (keV) & $E'$, (keV) & $\mathrm{\varepsilon (\%)}$ & $N$\\
     \hline
        Low &0.6-500     &0.6-495       &85\tnote{b} & 19005 \\
       Medium &500-4000    &511-2800      &98.1 & 25091 \\  
      High &4000-7000 &4000-7000\tnote{a} &86.5 & 30590\\
    \hline
     \end{tabular*}
     \begin{tablenotes}
     \item[a] The high-energy reference data set did not require rescaling.
     \item[b] The signal survival probability is energy-independent only above $92.36 \:\mathrm{keV}$. For energy dependence at lower energies (see Fig.~\ref{fig:signal_survival}).
     \end{tablenotes}
     \end{threeparttable}
      \end{table}
The lower energy limit of the low-energy data set is the threshold of the TUM40 detector module of $0.6\,\mathrm{keV}$ \cite{Angloher:2014myn}, whereas the upper limit marks the point at which saturation effects cannot be compensated reliably anymore by the applied energy reconstruction. With respect to the DM analysis \cite{florian} of this data set we modified the energy limits: We expand the upper limit to $\sim 500\,\mathrm{keV}$ by loosening the cuts a bit. However, the lower limit we increase as below $1\,\mathrm{keV}$ events occur of unknown origin. As their modelling is beyond the scope of this work, we disregard the events in the $0.6-1\,\mathrm{keV}$ range. Hence, hereafter ROI refers to the energy range $1-40\,\mathrm{keV}$. Within this ROI we count \num{11145} background events in TUM40, resulting in the total net rate $R_\mathrm{TUM40}=2.2\,\mathrm{kg^{-1}\, keV^{-1}\, d^{-1}}$ using the signal survival probability $\mathrm{\varepsilon}$ (see Fig.~\ref{fig:signal_survival}).\par
The energy scale of the reference data set agrees well with the literature values of identified peaks for low energies, i.e. in energy range relevant for DM search that is  $0.6-40\,\mathrm{keV}$ for TUM40 \cite{Angloher:2014myn}. For energies above $\sim 250\,\mathrm{keV}$, the energy scale starts to deviate by more than $1\,\mathrm{keV}$ as expected due to the non-linearities \cite{dissertation_cenk}. To correct this shift in the low- and medium-energy data sets, a rescaling is performed via a cubic polynomial. The values of the free parameters are determined by minimising the deviation between the measured energy and the literature value of identified peaks (see \cite{dissertation_cenk} for details).\par
\begin{figure}[htbp]
\subfloat[]{
	\label{fig:spectrum_low}
	\includegraphics[width=\linewidth]{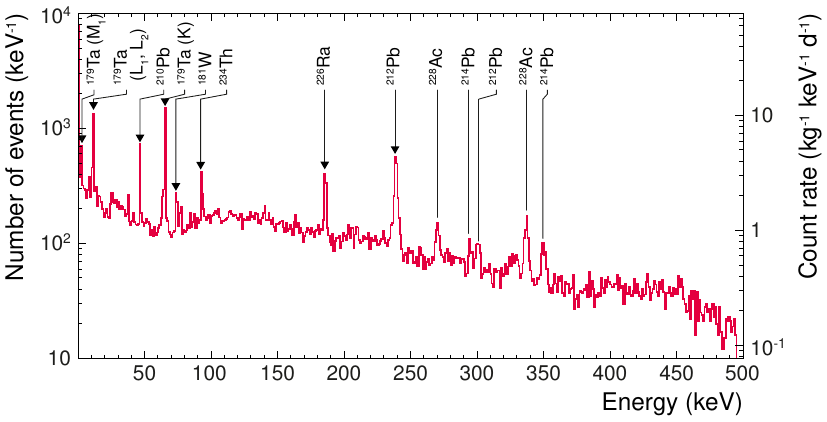}
}\\
\subfloat[]{
	\includegraphics[width=\linewidth]{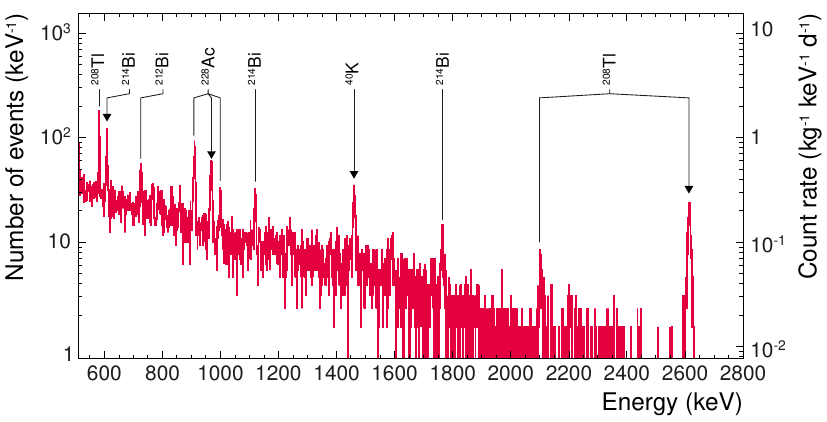}
	\label{fig:spectrum_medium}
}\\
\subfloat[]{
	\includegraphics[width=\linewidth]{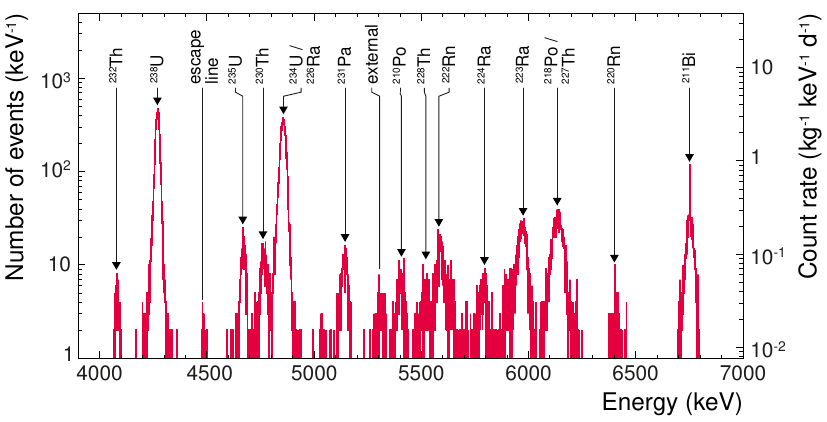}
	\label{fig:spectrum_high}
}
\caption{Reference data after rescaling for \protect\subref{fig:spectrum_low} the low-energy ($0.6\,\mathrm{keV}$ to $495\,\mathrm{keV}$), \protect\subref{fig:spectrum_medium} the medium-energy ($511\,\mathrm{keV}$ to $2800\,\mathrm{keV}$) and \protect\subref{fig:spectrum_high} the high-energy ($4000\,\mathrm{keV}$ to $7000\,\mathrm{keV}$) range. For all three histograms the bin size is $1\,\mathrm{keV}$. Triangles indicate peaks used for the normalisation of the simulation in this work, for details see text.
}
\label{fig:refdata}
\end{figure}

The high-energy reference data set includes sharp peaks which originate from $\alpha$-decays within the $\mathrm{{}^{232}Th}$, $\mathrm{{}^{235}U}$ and $\mathrm{{}^{238}U}$ decay chains (see Fig.~\ref{fig:spectrum_high}) mainly as bulk contamination. Besides the full absorption lines, also an escape peak around $4.5\,\mathrm{MeV}$ due to $\mathrm{{}^{230}Th}$ ($Q=4769.8\,\mathrm{keV}$, $E_\gamma=253.7\,\mathrm{keV}$) and $\mathrm{{}^{235}U}$ ($Q=4678\,\mathrm{keV}$, $E_\gamma=185.2\,\mathrm{keV}$) decays is visible, labelled as \textit{escape line}. In addition to bulk contamination, also a peak due to near-surface decay of $\mathrm{^{210}Po}$ is visible, labelled as \textit{external}. Due to the applied cuts, pile-up events caused by the correlated decays $\mathrm{^{212}Bi} \rightarrow \mathrm{^{212}Po}$, $\mathrm{^{214}Bi} \rightarrow \mathrm{^{214}Po}$, and $\mathrm{^{219}Rn} \rightarrow \mathrm{^{215}Po}$, which were observed in \cite{Strauss:2014aqw,Munster:2014mga}, are not included in the reference data. Because of the non-linearity of the TES, the calibration was corrected by assigning Q-values from literature \cite{iaea} to well identified peaks and interpolate linearly between them. The spike in the middle of the $\mathrm{{}^{211}Bi}$ peak (see Fig.~\ref{fig:spectrum_high}) is an artefact of the energy calibration\footnote{This is because the "fixpoint" of the polynomial, which is used to map the measured voltage signals to calibrated energies, is the centre of the reference peak. Therefore, events close to this energy deviate less and the peak gets "compressed".} of the data.\par
As the electromagnetic background components clearly dominate the total background \cite{Angloher:2011uu,Angloher:2014myn, Strauss:2014aqw}, we do not take into account e.g.\ neutron induced nuclear recoils as background. Therefore, no differentiation between the $\beta$/$\gamma$-band and the nuclear recoil bands is needed. Hence, we neglect the light yield information in the reference data set and will compare our background model to energy spectra of the total background.
The final spectra for the three reference data sets are shown in Fig.~\ref{fig:refdata}. The medium and high-energy data sets are used to normalise the background model which will be described in Section~\ref{sec:4}. The low-energy data set is used to cross check the accuracy of the model and to determine the background contribution in the ROI which will be discussed in Section~\ref{sec:results}.

\section{Simulation of the background spectra} \label{montecarlosim}
For an accurate simulation of the background spectra, hereafter called \textit{spectral templates of individual decays}, we use a two-stage approach: for the microscopic simulation of the energy deposition in the detector parts we use a Geant4-based simulation code. To apply the detector response we process the simulated data with a ROOT-based tool \cite{Brun:1997pa}. This approach has the advantage that changes in the detector parameters do not require to rerun the often time-consuming Geant4 simulations. \par
Furthermore, as we exclude pile-up events from the experimental data (see Section~\ref{reference_data}), we are allowed to split decay chains $X_1 \rightarrow X_2 \rightarrow X_3 \rightarrow \ldots$ into piece-wise decays $X_1 \rightarrow X_2$, $X_2 \rightarrow X_3$, \ldots This has the advantage that the activities $A_1$, $A_2$, \ldots associated with nuclides $X_1$, $X_2$, \ldots stay free and can be normalised to calibration or assay measurements without the need to rerun the simulations (see Section~\ref{sec:norm}).

\subsection{Implementation of detector geometry and physics} \label{geant4}
For this work we greatly extended the simulation tool \textit{ImpCRESST} \cite{scholl2011} and adapted it for Geant4 version 10.2 patch 1 to simulate the decay of the primary contaminants, the production of the resulting secondary particles, the tracking of all particles through a detector geometry, and the energy deposition in the instrumented detector parts. To obtain reliable simulation results we adjusted the physics list to sub-keV energies, implemented a detailed geometry of our detector, and verified the decay properties of the contaminants as primary input to the simulation. \par
We use the physics list \textit{Shielding} as provided by Geant4 with the following three modifications: \par
(i) Instead of the standard physics constructor \texttt{G4EmStan\-dardPhysics}, we chose \texttt{G4EmStandardPhysics\_option4} because it provides the most accurate modelling of electromagnetic interactions at low energies \cite{G4LowEnergy}.\par
(ii) As production cut for secondaries, we set a value of $250\,\mathrm{eV}$ throughout all volumes\footnote{This is the lowest value for which all relevant electromagnetic physics processes are approved. We note that for specialised use cases the applicability limit can be lower, e.g.\ for elastic electron scattering in silicon as low as $5\,\mathrm{eV}$. By default the cut value is $990\,\mathrm{eV}$ \cite{G4LowEnergy}.}. If the kinetic energy of a potential secondary drops below this value, no actual secondary is simulated but energy conservation is obeyed by locally depositing the energy. By choosing this low value for the production cut, we consider the leakage of radiation out of the finite detector volume via secondary radiation to a high accuracy. Regardless of this production cut, secondary particles caused by atomic de-excitation, e.g.\ fluorescence photons and Auger electrons, are produced in any case. \par
(iii) Radioactive decays are handled by the \texttt{G4Radio\-activeDecayPhysics} module. As described in \cite{G4Tritium}, we patched the code to correctly treat $\mathrm{{}^{3}H}$ as unstable. We note that for a proper simulation of the $\mathrm{{}^{234}Th}$-decay, the most prominent internal background source for the target crystal, at least Geant4 version 10.2 is needed.\par
\begin{figure}[htbp]
\includegraphics[width=\linewidth]{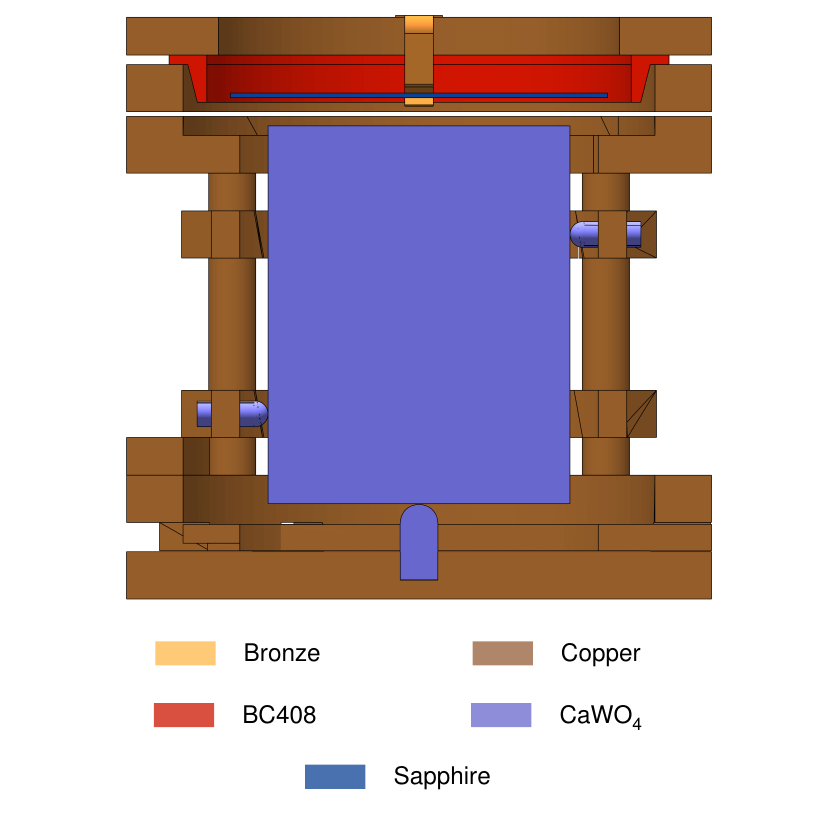}
\caption{Cut-away view parallel to the $X$-$Z$-plane of the TUM40 detector module as implemented in Geant4.}
\label{fig:tum40_geant}
\end{figure}
The modelled geometry is shown in Fig.~\ref{fig:tum40_geant} and reflects closely the shape and size of the actual detector module TUM40 (see Fig.~\ref{fig:tum_photo}). In particular, we considered the block-shaped target crystal made of $\mathrm{CaWO_4}$ which is held by sticks of the same material and copper parts. The target is faced by the light detector which is approximated as pure sapphire ($\mathrm{Al_2O_3}$), contrary to the SOS used in reality. We omit the silicon layer of $1\,\mathrm{\upmu m}$-thickness on the back side of the light detector in the simulation because of its negligible mass contribution. With the same argument we omit both TESs, the wires, and the bronze clamps from the outside pushing the $\mathrm{CaWO_4}$ sticks. \par
Both the target crystal and the light detector are surrounded by the scintillating foil, which we approximate with Mylar ($\mathrm{C_{10}H_{8}O_{4}}$) foil in the simulation. We consider the ring holding the foil, which is made out of plastic scintillator BC408, and the bronze clamps, which hold the light detector. The properties of the simulated materials are listed in Table~\ref{tab:materials}. For all elements, we assume a natural isotopic abundance.\par
For this work, we only record the Geant4 \textit{hits} in the target crystal for which we store the deposited energy $E_\mathrm{dep}$ and the time information $t$.
\begin{table}
\centering
\caption{Chemical composition, total mass per detector part, and material density as implemented in the simulation.}
\label{tab:materials}
\begin{threeparttable}
\begin{tabular*}{\columnwidth}{@{\extracolsep{\fill}}lcS[table-format=3.1(0)]S[table-format=1.2(0)]@{}}
\hline
Detector part      & Composition & {Mass (g)} & {Density ($\mathrm{g\,cm^{-3}}$)}\\
\hline
Copper parts       & Cu                      & 469.2 & 8.92\\
Target crystal     & $\mathrm{CaWO_4}$       & 246.2 & 6.01\\
Holder sticks      & $\mathrm{CaWO_4}$       & 2.5 & 6.01\\
Light detector     & $\mathrm{Al_2O_3}$      & 2.5 & 3.99\\
Scintillating foil & $\mathrm{C_{10}H_8O_4}$ & 1.3 & 1.37\\
BC408 ring         & $\mathrm{C_{10}H_{11}}$ & 1.3 & 1.03\\
Bronze clamps\tnote{a}      & 6\%\,Sn + 94\%\,Cu         & 0.4 & 8.82\\
\hline
\end{tabular*}
\begin{tablenotes}
     \item[a] Composition given by mass fractions.
\end{tablenotes}
\end{threeparttable}
\end{table}

\subsection{Emulation of the detector response}
To emulate the finite time and energy resolution of the detector, we apply an empirical detector response model to the outcome of the Geant4 simulation.
To consider the time resolution of the detector, we sum up all Geant4 hits which happen within $\Delta t$ with respect to the first hit of a simulated decay to create one \textit{experimental event} with energy $E_\mathrm{dep}$.
If hits happen after $t + \Delta t$, e.g.\ via a delayed decay of an isomeric state, the next experimental event will be immediately created. This procedure is applied until all detector hits are processed. We found that $\Delta t=2\,\mathrm{ms}$ describes the experimental data well. We note that this is a purely empirical approach and does not trivially correspond to length, rise time, dead time etc. of the experimental detector pulses. As described in Section~\ref{reference_data}, the reference data contains none of the correlated decays, hence also the simulation omits them. \par
\begin{figure}[htbp]
\includegraphics[width=\linewidth]{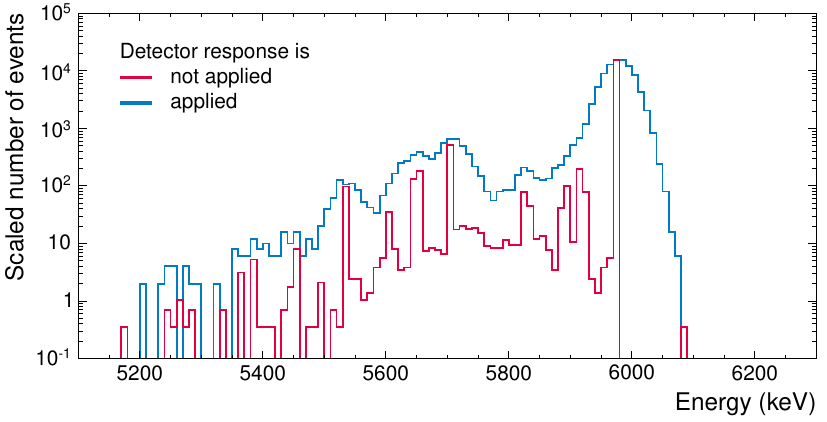}
\caption{Simulated energy deposition inside the $\mathrm{CaWO_4}$ target crystal caused by a bulk contamination with $\mathrm{{}^{223}Ra}$ without (\textit{red} histogram) and with (\textit{blue} histogram) applied detector response model. The maximum of latter histogram is scaled to the maximum of former; bin size is $10\,\mathrm{keV}$}
\label{fig:resp_model}
\end{figure}
For each energy deposition $E_\mathrm{dep}$ of an event, we calculate the observable energy $E_\mathrm{obs}$ by randomly sampling a Gaussian $G(E_\mathrm{obs};E_\mathrm{dep}, \sigma(E_\mathrm{dep}))$ centred at $E_\mathrm{dep}$. The energy-dependent variance was obtained from fitting cubic polynomials to the experimental detector resolution via reference data sets in
the low ($\sigma_\mathrm{l}$), medium ($\sigma_\mathrm{m}$), and high ($\sigma_\mathrm{h}$) energy range (for details see \cite{dissertation_cenk}). The effect of the finite energy resolution is clearly shown on the example of the $\mathrm{{}^{223}Rn}$ decay spectrum in Fig.~\ref{fig:resp_model}.

\subsection{\label{sec:norm}Simulation of spectral templates}
A \textit{spectral template} $T_{v i X}(E_\mathrm{obs})$ is a probability density function (PDF) which gives the probability to observe an experimental event with energy $E_\mathrm{obs}$ in the instrumented volume $i$, caused by a bulk contamination $X$ in volume $v$. Hence, it has to be scaled to the contamination level to give the total number of background events. We obtain it in five steps:
(i) we sample the volume $v$ homogeneously $N_{v,X}$-times, each time placing a nuclide of type $X$ at rest; (ii) in \textit{ImpCRESST} the decay $X \rightarrow Y$ of the nuclide is simulated; (iii) the detector response model is applied; (iv) we count how many of the decayed nuclides lead to an energy deposition $E_\mathrm{obs}$ in volume $i$: $n_i(E_\mathrm{obs})$; (v) we obtain $T_{v iX}(E_\mathrm{obs})=n_i(E_\mathrm{obs})/N_{v,X}$.
Assuming that the counting of $n_i(E_\mathrm{obs})$ is a Poisson process we define its statistical uncertainty as $\sqrt{n_i(E_\mathrm{obs})}$ and propagate it for all the dependent quantities.\par
\begin{figure}[htb]
\includegraphics[width=\linewidth]{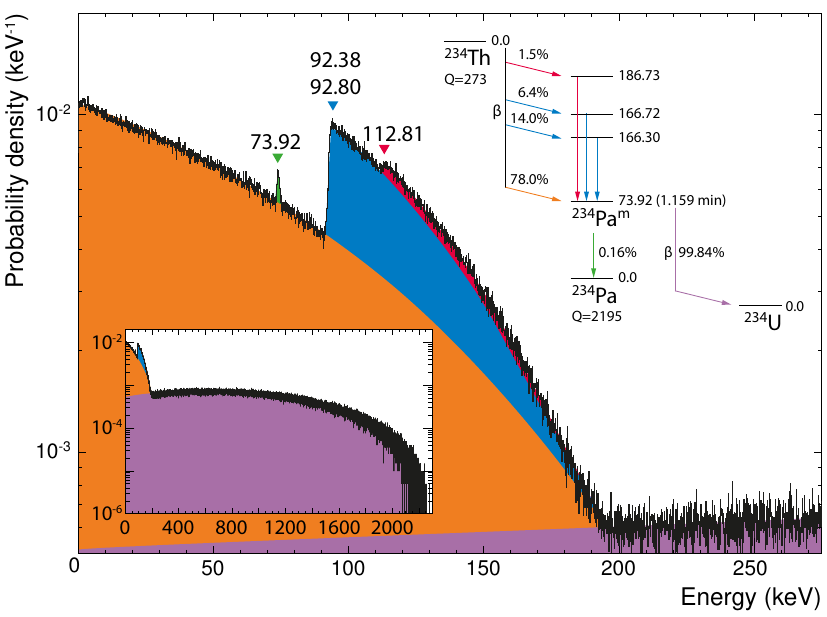}
\caption{Simulated decay of $\mathrm{{}^{234}Th}$ with applied detector response model. Due to the isomeric state $\mathrm{{}^{234}Pa^m[73.92\,keV]}$ the decay results in prompt (\textit{orange}, \textit{blue}, and \textit{red} filled histograms) and delayed (\textit{violet} filled histogram and \textit{green} peak at $\mathrm{73.92\,keV}$) detector events. The bin size is $100\,\mathrm{eV}$. For details see text; data for the level scheme are taken from \cite{iaea}.}
\label{fig:234Th}
\end{figure}
As an example, Fig.~\ref{fig:234Th} shows the sum of the two spectral templates which describe the $\beta$-decay of $\mathrm{{}^{234}Th}$: 
either directly or indirectly, the decay reaches the isomeric state $\mathrm{{}^{234}Pa^m[73.92\,keV]}$ (cf.\ the level scheme in Fig.~\ref{fig:234Th}). Due to the long half-life of this state ($T_{1/2}=1.159\,\mathrm{min}$ \cite{iaea}) compared to the time resolution of the detector $\mathcal{O}(\mathrm{ms})$, the decay cascade results in two detector events: one containing the interactions before reaching the isomeric state and one afterwards. \par
The isomeric state can be reached via four branches: once via a direct $\beta$-decay (\textit{orange} filled histogram in Fig.~\ref{fig:234Th}) and three times via a $\beta$-decay to a state at energies above the isomeric state with a subsequent $\gamma$-decay to the isomeric state (\textit{blue} and \textit{red} filled histograms in Fig.~\ref{fig:234Th}). Due to the slow detector response, the $\gamma$-decay cannot be resolved and the pile-up results in a $\beta$-spectrum shifted to the energy of the associated $\gamma$-line as obvious for the \textit{blue} and \textit{red} filled histograms. \par
In 0.16\,\% of the cases the $\mathrm{{}^{234}Pa^m[73.92\,keV]}$ state decays to the ground-state of $\mathrm{{}^{234}Pa}$ \footnote{We note that prior to version 10.2, Geant4 did not implement this decay properly \cite{G4234Th}: $\mathrm{{}^{234}Pa^m[73.92\,keV]}$ decayed in 100\,\% of the cases to the ground state of $\mathrm{{}^{234}Pa}$, resulting in an overestimation of the monochromatic line at $73.92\,\mathrm{keV}$.}, resulting in a monochromatic line at $73.92\,\mathrm{keV}$ as the second detector event \cite{iaea} (\textit{green} peak in Fig.~\ref{fig:234Th}). In the remaining 99.84\,\% of the cases, the isomeric state undergoes a $\beta$-decay to the ground state of $\mathrm{{}^{234}U}$ \cite{iaea} resulting in a $\beta$-spectrum with an endpoint of $2.27\,\mathrm{MeV}$ as the second detector event (\textit{violet} filled histogram in Fig.~\ref{fig:234Th}). \par
\begin{figure}[ht]
\includegraphics[width=\linewidth]{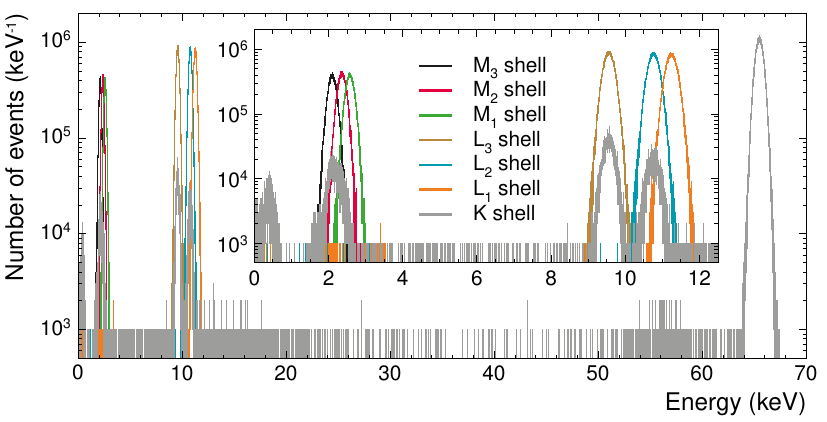}
\caption{Simulated energy deposition in $\mathrm{CaWO_4}$ caused by EC of $\mathrm{{}^{179}Ta}$ separated according to the involved atomic shells. The inset shows a zoom to the EC-peaks for the M- and L-shells. The bin size is $1\,\mathrm{eV}$.}
\label{fig:179Ta}
\end{figure}

As described in Section~\ref{sec:cosmicActivation} one cosmogenic background source is the EC decay of $\mathrm{{}^{179}Ta}$. However, we found that Geant4 assigns the same capture probability regardless from which atomic shells the electron is captured contrary to e.g.\ \cite{Firestone1996}. To mitigate this contradiction, we disentangle the spectral templates for the individual atomic shells and determine their activity from a fit to the reference data (see Table~\ref{tab:cosmogenic_activities} for the resulting activities), i.e.\ we do not rely on the contradicting literature values by leaving the capture probability unconstrained. We disentangle the cumulative decay spectrum, as simulated by Geant4, into individual components by selecting the events according to the total energy of all decay products including the recoiling $\mathrm{{}^{179}Hf}$ nucleus. Fig.~\ref{fig:179Ta} shows the resulting spectral templates, i.e.\ each histogram is normalised to one. We note that each atomic shell not only leads to a peak at the fully absorbed electron binding energy of $\mathrm{{}^{179}Hf}$, but may also causes escape peaks. This is clearly visible for the K-shell (Fig.~\ref{fig:179Ta}, \textit{grey} histogram) with a peak caused by the fully absorbed binding energy of  $65.35\,\mathrm{keV}$ \cite{Firestone1996} and escape peaks nearly coinciding with peaks of the fully absorbed binding energies of the $\mathrm{L_2}$, $\mathrm{L_3}$, and $\mathrm{M_3}$-shells.

\section{Normalisation of background contributions} \label{sec:4}
The next step is to scale these resulting templates according to the experimental reference data. After the scaling, these individual contributions are combined and compared to the reference data.

\subsection{Determination of the internal radiogenic background components} \label{internal_radiogenic}
Since it is complicated to disentangle each $\beta$-decay or Compton continuum as they overlap with each other, we correlate these spectra in the ROI to clearly defined $\alpha$ lines in the energy range between $4\,\mathrm{MeV}$ and $7\,\mathrm{MeV}$. The connection between the $\alpha$ lines in the MeV range and the $ \beta$/$\gamma$-decays in the keV range comes from \textit{secular equilibrium} \cite{LAnnunziata2012xxxvii}. We omit the $\alpha$ lines of $\mathrm{^{147}Sm}$ and $\mathrm{^{180}W}$ which were observed in an earlier analysis \cite{Strauss:2014aqw} as they are not connected to decays in the ROI.\par
In this work, we use two requirements for the existence of secular equilibrium: (i) the relative difference of the activities of the parent $A$ and the daughter $B$ has to be $(A_{A}-A_{B})/A_{A} < 1 \, \%$ after 460 days had passed, which is the time elapsed between the crystal growth and its arrival underground. (ii) We assume that the initial activity $A_{B}(\mathrm{time}=0)$ of the daughter nuclide is negligible if it decays to less than 10\,\%  over time before measurement. If both of these requirements are fulfilled, we consider the parent and the daughter to be in secular equilibrium, so their activities are the same, otherwise, their activities are determined independently.\par
After identifying the $\alpha$ peaks in the high-energy reference data (see Section~\ref{reference_data}), we determine the number of events within a given peak $X$ of amplitude $\tilde{A}$ and width $\sigma$ by fitting a non-normalised Gaussian. Hence, the total number of observed events $N_{X,\mathrm{obs}}$ of the $\alpha$ line $X$ is $\tilde{A} \cdot \exp(-E'^2/(2 \sigma^2))$. We consider the fit uncertainty of $\tilde{A}$ and $\sigma$ as a systematic uncertainty with respect to this MC study and propagate it as absolute systematic uncertainty for all the dependent quantities. We ignore any potential correlations between the fitted parameters $\tilde{A}$ and $\sigma$. For the total uncertainty we linearly combine this systematic uncertainty with the previous introduced statistical uncertainty of the simulated templates. From the total number of observed events, the activity of the $\alpha$ line $X$ is calculated as:
\begin{equation}
A_{X} = \frac{N_{X, \: \mathrm{obs}}}{\mathrm{\varepsilon_h} \cdot T \cdot M} \label{alpha_activity},
\end{equation}
where $T \cdot M$ stands for the exposure, $M$ is the mass of the $\mathrm{CaWO_{4}}$ crystal (see Table \ref{tab:materials}), $T$ is the live-time and $\varepsilon_{\mathrm{h}}$ for the signal survival probability of the high-energy reference data set (see Section \ref{reference_data}). The obtained activities are listed  together with the fit parameters in Table~\ref{tab:alpha_activities} in \ref{appendix}.\par
\begin{figure}[htbp]
\includegraphics[width=\linewidth]{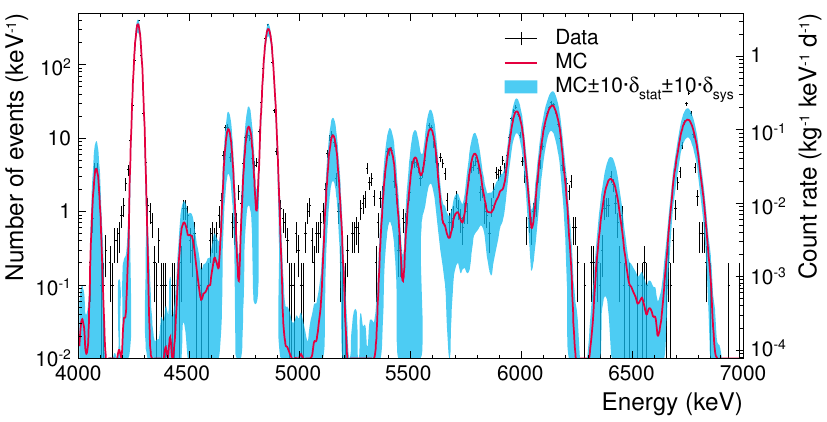}
\caption{Comparison of simulated $\alpha$-spectrum (\textit{red}, without the external $\mathrm{^{210}Po}$ contamination at $5.3\,\mathrm{MeV}$) and the experimental energy spectrum of TUM40 (\textit{black}). \textit{Light blue} shows the band of statistical and systematic uncertainties multiplied by a factor 10 for better visibility. The bin size is $10\,\mathrm{keV}$.}
\label{fig:alfa}
\end{figure}

To compare the simulation results with the reference data (see Fig.~\ref{fig:spectrum_high}), the templates of the individual $\alpha$-decays were scaled with their respective activities and combined (see Fig.~\ref{fig:alfa}). The resulting ratio of the integral of counts in the simulated spectrum to the counts in the experimental data between $4\,\mathrm{MeV}$ and $7\,\mathrm{MeV}$ is $(97.8 \pm 3.8) \, \%$. The uncertainty band shown in Fig.~\ref{fig:alfa} includes both statistical and systematic uncertainties from the simulated templates and the fit of the reference data, respectively.
The next step is to determine the activities $A_{Y, \: \mathrm{pred} \: (\beta)}$ of the $\beta$/$\gamma$-decaying nuclides $Y$, which are in secular equilibrium with their $\alpha$-decaying parents $X$:
\begin{equation}  \label{beta_activity}
A_{Y, \: \mathrm{pred} \: (\beta)} = A_X \cdot B.R._{X \to Y}, 
\end{equation}
where $B.R._{X \to Y}$ is the cumulative branching ratio between the nuclides $X$ and $Y$.
For the comparison between the predicted count spectrum and the observed one in Section~\ref{sec:results} we also calculate the number of observable predicted events $N_{Y, \: \mathrm{pred} \: (\beta)}$ as:
\begin{equation}
N_{Y, \: \mathrm{pred} \: (\beta)} = A_{Y, \: \mathrm{pred} \: (\beta)} \cdot \mathrm{\varepsilon}_{\mathrm{l}} \cdot T \cdot M. \\
\label{beta_events}
\end{equation}
The breakdown of the decay chains with respect to the piece-wise secular equilibrium are given in \ref{appendix} in Tables \ref{tab:U238}, \ref{tab:U235} and \ref{tab:Th232}. The tables list also the activities as determined for the $\mathrm{CaWO_4}$ crystal of TUM40.

\subsection{Near external radiogenic background components} \label{near_external_radiogenic}
The main source of near external radiogenic contamination is expected from the copper holders of the detector modules due to its relatively large mass close to the crystal. This type of background manifests itself in the ROI as $ \beta$/$\gamma$-decays. The copper holders are separated from the crystal by a thin scintillating foil, so even external low-energy $\beta$s may be visible in the experimental data. The detector holder of TUM40 is made of NOSV copper from Norddeutsche Affinerie AG (see Section~\ref{ambient_gamma}), the same copper as used in the CUORE experiment \cite{Alduino:2016vjd}. Under the assumption that both experiments use copper from the same batch \cite{dissertation_cenk} we use the upper limits on the bulk activities from \cite{Alduino:2016vjd}: $A_{\mathrm{Cu, \: ^{238}U}} \leq 65 \: \mathrm{\upmu Bq \, kg^{-1}}$, $A_{\mathrm{Cu, \: ^{232}Th}} \leq 2 \: \mathrm{\upmu Bq \, kg^{-1}}$.
We found no literature values for the specific activity of $\mathrm{{}^{235}U}$ in copper. As suggested in \cite{APRILE201143}, we assume the same isotopic abundance as in natural uranium and derive the upper limit: $A_{\mathrm{Cu, \: ^{235}U}} = A_{\mathrm{Cu, \: ^{238}U}} \cdot 0.70 \,\% \leq 0.46\,\mathrm{\upmu Bq \, kg^{-1}}$.
Because only activities for the nuclides at the head of the corresponding decay chain are reported in \cite{Alduino:2016vjd} we assume that the chains are in secular equilibrium to obtain the activities for the daughter nuclides. This assumption is supported by measurements of the XENON experiment for another copper batch from the same producer \cite{APRILE201143}.\par
The mass determines the activity whereas the placement determines what fraction of the activity is visible to the target crystal as background. So, instead of a time-consuming simulation of all contaminants in all copper parts, we simulate the contaminants only in the copper part which has the highest visibility but scale the obtained background spectra to the total mass of all near copper parts. Considering this information, the scaling of the individual decay templates is performed as follows:
\begin{equation} \label{near_external_beta_eq}
N_{X, \: \mathrm{pred}} = A_\mathrm{top} \cdot B.R._{\mathrm{top} \to X} \cdot \mathrm{\varepsilon_{\mathrm{l}}} \cdot T \cdot m_{\mathrm{Cu}},
\end{equation}
where $A_{\mathrm{top}}$ stands for the activity of the head of the corresponding decay chain, $B.R._{\mathrm{top} \to X}$ stands for the cumulative branching ratio between the head of the decay chain and the nuclide $X$, and $m_{\mathrm{Cu}}$  stands for the total copper mass (see Table \ref{tab:materials}).
As we use upper limits for the activities to scale the spectral templates, no systematic uncertainties are taken into account; only the statistical uncertainties associated with the simulation of the templates are used.

\subsection{Additional external radiogenic background components} \label{distant_external_radiogenic}
After removing the contributions of internal radiogenic and near external radiogenic contributions from the experimental reference data, we still observe $\gamma$-peaks. These identified lines belong to the decays of nine radioactive nuclides: $\mathrm{^{40}K}$, $\mathrm{^{208}Tl}$, $\mathrm{^{210}Pb}$, $\mathrm{^{212}Bi}$, $\mathrm{^{212}Pb}$, $\mathrm{^{214}Bi}$, $\mathrm{^{226}Ra}$, $\mathrm{^{228}Ac}$, and $\mathrm{^{234}Th}$. Hence, at least for these nine nuclides the total activity of the near copper parts is not enough to explain the observation. It is reasonable to assume that this additional background is caused by parts of the detector modules or shielding structure which have not yet been considered as a source of background in our simulations: massive but more distant copper parts (e.g.\ thermal shieldings of the cryostat) or/and closer but less massive components of the detector module (e.g.\ scintillating foil, SOS, TES, wires, bronze clamps). As simulating the complete decay chains in the full experimental setup was beyond the scope of this work, we did a first approximation of the effects of these additional background components. As copper is the most abundant material close to the detectors, we created as a first approximation a $1\,\mathrm{mm}$-thick spherical copper shell surrounding the detector module and simulated only these nine clearly identified nuclides as its bulk contaminants. The shell is therefore a very simplified model of the distant copper parts. We note that in reality this background may originate partially also in the other mentioned components; it is up to a future study to identify them.\par
A consequence of this simplified model is that we do not know the geometrical efficiency, i.e.\ the probability for a background particle originating in some of these additional components to reach the target crystal. Hence, we can only give the activity $A$ that the crystal would be able to observe if survival probability of the particle was 100 \%.
The determination of the activity corresponding to the observed $\gamma$-lines is very similar to the procedure outlined in Section~\ref{internal_radiogenic} for the determination of the activity, i.e.\ it is the background rate corrected for the signal survival probability. However, there exists a continuous spectrum in the low and medium energy ranges which we consider as a linear component of the fit function. In order to get the number of events corresponding only to the $\gamma$-peaks, we calculate the number of events $N_{\mathrm{peak}, \: \mathrm{obs}}$ below the Gaussian component of the fit.
Consequently, the activity $A$ corresponding to the $\gamma$-lines can be calculated using Eq.~\ref{alpha_activity}. 
In order to calculate the activity of the full decay, one has to scale the activity of the $\gamma$-peak with $\eta$,
which is the fraction of events under the $\gamma$-peak compared to all the events of the particular decay as calculated from the simulation templates. The total activity of each isotope is then $A_\mathrm{total, \: obs} = A_\mathrm{peak, \: obs}/\eta$.
The activity $A_\mathrm{total, \: obs}$ of the $\gamma$-lines, the fraction $\eta$ and the corresponding decays are listed in Table \ref{tab:gamma_activities} in \ref{appendix}. In order to predict the observable number of events $N_\mathrm{total, \: pred}$ in the low-energy range, we use Eq.~\ref{beta_events}.

\subsection{Determination of internal cosmogenic background components} \label{internal_cosmogenic}
Internal cosmogenic background components contribute to the overall background in the ROI in the form of $\beta$-decay (e.g.\ $^{3}\mathrm{H}$) or in the form of X-ray and Auger electron emissions that result from EC (e.g.\ $^{179}\mathrm{Ta}$ and $^{181}\mathrm{W}$). \par
After removing the internal and external (both near and additional) radiogenic background components, the activities of $^{179}\mathrm{Ta}$ and $^{181}\mathrm{W}$ nuclides are determined in the same way as the activities of the external $\gamma$-rays (see Section~\ref{distant_external_radiogenic}).
$^{3}\mathrm{H}$ may be produced via cosmic activation in several ways. However, the only reaction path which is empirically evident for TUM40 is the co-production together with $^{181}\mathrm{W}$ (see Eq.~\ref{183W}). Hence, we use the activity of $^{181}\mathrm{W}$ to estimate a lower limit on the $^{3}\mathrm{H}$ contamination. A first calculation using ACTIVIA \cite{Back:2007kk} indicated that the activity of $^{3}\mathrm{H}$ is 62\,\% of the activity of $^{181}\mathrm{W}$.
The activities calculated for each peak and the corresponding nuclide are given in Table \ref{tab:cosmogenic_activities} in \ref{appendix}.

\section{Discussion of the results}\label{sec:results}
In this section, we first check the self-consistency of our normalisation procedure used in Section~\ref{sec:norm} by investigating the compatibility of the medium-energy reference data with the background model. Afterwards, we discuss the background prediction in the ROI and its agreement with the observation. The comparison of the combined model components with the experimental reference data sets are shown in Fig.~\ref{fig:result}.
\subsection{Self-consistency of the simulation}
\begin{figure*}[p]
\centering
\subfloat[]{
    \includegraphics[width=0.72\linewidth]{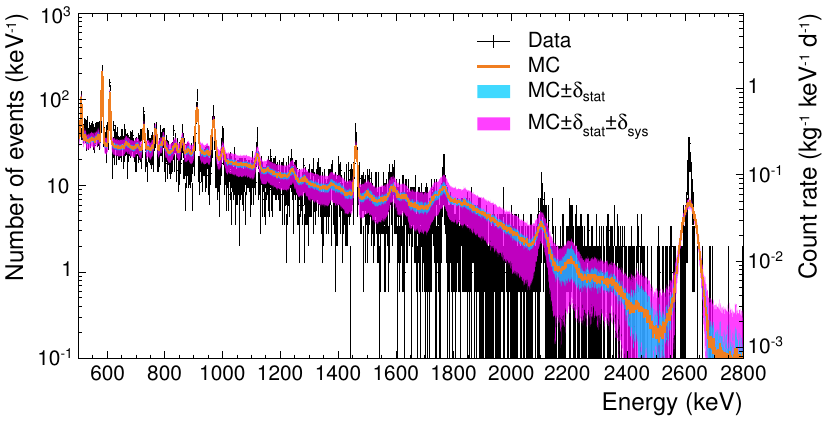}
    \label{fig:result_medium2800}
}\\
\subfloat[]{
    \includegraphics[width=0.72\linewidth]{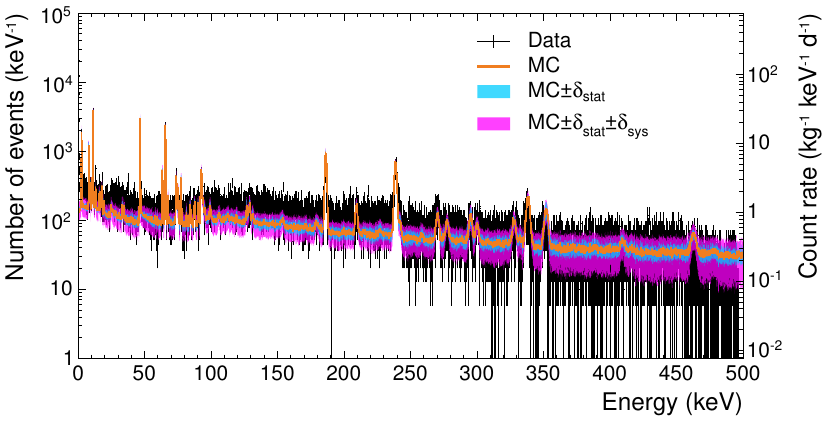}
    \label{fig:result_low500}
}\\
\subfloat[]{
    \includegraphics[width=0.72\linewidth]{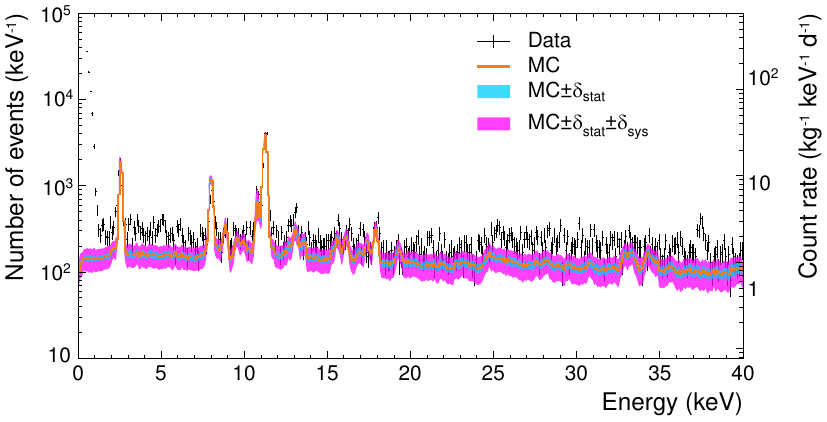}
    \label{fig:result_low40}
}
\caption{Total simulated background spectrum (MC, \textit{orange} histogram) with statistical (\textit{cyan} band) and systematic (\textit{magenta} band) uncertainties with respect to the experimental data with statistical uncertainties (\textit{black} data points) in the medium-energy range \protect\subref{fig:result_medium2800}, the low-energy range \protect\subref{fig:result_low500}, and the ROI $(1-40\,\mathrm{keV})$ \protect\subref{fig:result_low40}. The bin sizes are $1\,\mathrm{keV}$, $100\,\mathrm{eV}$, and $100\,\mathrm{eV}$, respectively.}
\label{fig:result}
\end{figure*}

Fig.~\ref{fig:result_medium2800} shows the comparison between medium-energy reference data and the combined templates. We obtain the predicted number of background events in the medium energy range $N_\mathrm{MC, \: m}$ by adapting Eqs.~\ref{beta_events}, \ref{near_external_beta_eq} for the discussed background components.
The continuous part of the simulated spectrum matches well with the experimental data within the uncertainties.
Besides the $\gamma$-lines used to normalise the templates (see Fig.~\ref{fig:refdata}), also other prominent $\gamma$-lines belonging to nuclides $\mathrm{^{40}K}$, $\mathrm{^{208}Tl}$, $\mathrm{^{214}Bi}$ and  $\mathrm{^{228}Ac}$ fit well with regards to amplitude and width. This shows the self-consistency of the simulated spectral templates.\par
To quantitatively compare the simulation with the reference data we define $\zeta = N_{\mathrm{MC}}/N_{\mathrm{exp}}$, where $N_{\mathrm{exp}}$ is the number of events in the respective reference data set and $N_{\mathrm{MC}}$ is the number of predicted background events. In the following we assume that the near external radiogenic background component reachs the maximum activity that is allowed by the upper limit on the copper contamination (see Section~\ref{near_external_radiogenic}).
In the medium energy range between $511\,\mathrm{keV}$ and $2800\,\mathrm{keV}$ the reproduction of the background reaches up to $\zeta = (109 \pm 69)\,\mathrm{\%}$, 
i.e.\ the model matches the observed background within the uncertainty\footnote{Including statistical uncertainties from the simulated spectral templates and systematic uncertainties from the fit to the reference data.}. The uncertainty is dominated by the limited statistics of the experimental reference data set at medium energies (see Table~\ref{tab:reference data}). This limitation can only be resolved by increasing the exposure of the reference data set and is hence outside the scope of this work.\par
The comparison between model and reference data for the low-energy range is shown in Fig.~\ref{fig:result_low500}.
The combined peak due to the  $\mathrm{K_{\alpha1}}$ ($8.048\,\mathrm{keV}$) and $\mathrm{K_{\alpha2}}$ ($8.028\,\mathrm{keV}$) fluorescence lines in copper \cite{Firestone1996} is clearly visible in the experimental data and appears in situ also in the simulated spectrum (see Fig.~\ref{fig:result_low40}).
\begin{table}
    \centering
    \caption{Mean energy and counts of the combined $\mathrm{K_{\alpha1}}$ and $\mathrm{K_{\alpha2}}$ fluorescence lines of copper in the experimental reference data and the simulation. Rounded to two significant digits on the uncertainty.}
    \label{tab:flu}
    \begin{tabular*}{\columnwidth}{@{\extracolsep{\fill}} l S[table-format=1.4(5)] S[table-format=3.1(2)]}
    \hline
          & {Mean energy (keV)} & Counts\\ \hline
         Reference data & 8.0457 \pm 0.0042  & 180.2 \pm 6.4\\
         Simulation     & 7.9977 \pm 0.0026  & 260.4 \pm 6.4\\
         \hline
    \end{tabular*}
\end{table}
As this peak occurs due to the interaction of $\beta$/$\gamma$-background with the copper, one can consider it as an indicator of the correct estimation of the amount of external contamination. We determine the counts in this peak as described in Section~\ref{distant_external_radiogenic} and give the results in Table~\ref{tab:flu}. Despite the strong simplification of the model used for the distant copper parts, the simulation agrees with the reference data better than a factor two. \par
Even though the simulation and the experimental data agree within the uncertainties, it can be seen that the agreement decreases with decreasing energy. This means that we are either missing one or more $ \beta$/$\gamma$-decaying nuclides and/or we are underestimating one or more nuclides considered in this work. If we assume that the match between the simulation and the experimental data in the medium-energy range is correct, then the Q-value of the missing nuclide(s) in the low-energy range should be lower than $495\, \mathrm{keV}$, which is the upper limit of this energy range.\par
\subsection{Background reproduction at low energies}
The coverage of the low-energy reference data by the simulation is up to $\zeta  = (79 \pm 33)\,\mathrm{\%}$ (see Fig.~\ref{fig:result_low500}).
The comparison for the energy range between $1\,\mathrm{keV}$ and $40\,\mathrm{keV}$ (ROI) (see Fig.~\ref{fig:result_low40}), results in up to  $\mathbf{\zeta = (68 \pm 16)\,\mathrm{\%}}$.
The reduced uncertainty, going down from $33\,\mathrm{\%}$ to $16\,\mathrm{\%}$, is caused by the increasing statistics in the experimental reference data sets at lower energies. Higher statistics result in smaller uncertainties in the fit of the model components. The decrease of the fit uncertainty, as part of the systematic uncertainty, with decreasing energy is also clearly shown in Figs.~\ref{fig:result_low500}, \ref{fig:result_low40}.\par
\begin{figure*}[p]
\centering
\subfloat[]{
	\includegraphics[width=0.72\linewidth]{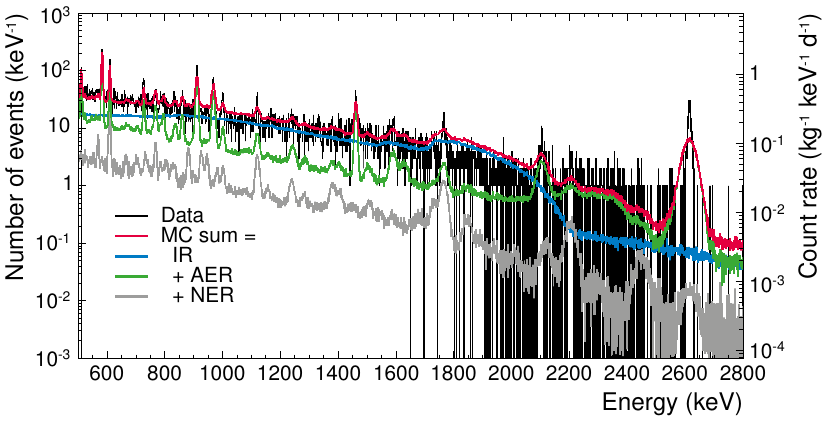}
	\label{fig:breakdown_2800}
}\\
\subfloat[]{
	\includegraphics[width=0.72\linewidth]{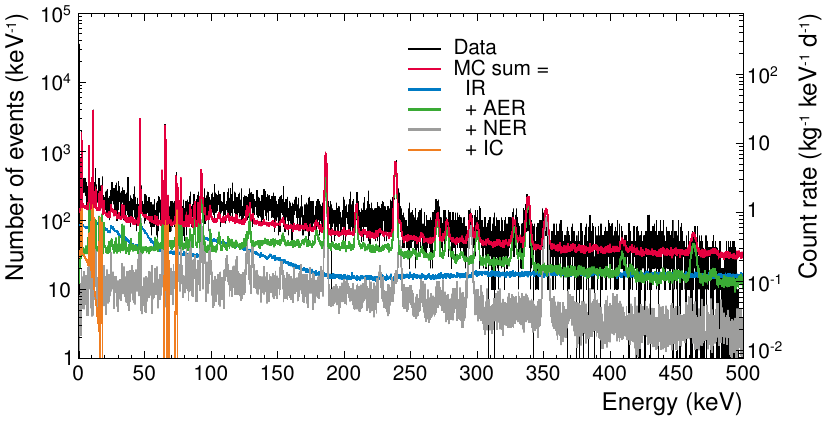}
	\label{fig:breakdown_500}
}\\
\subfloat[]{
	\includegraphics[width=0.72\linewidth]{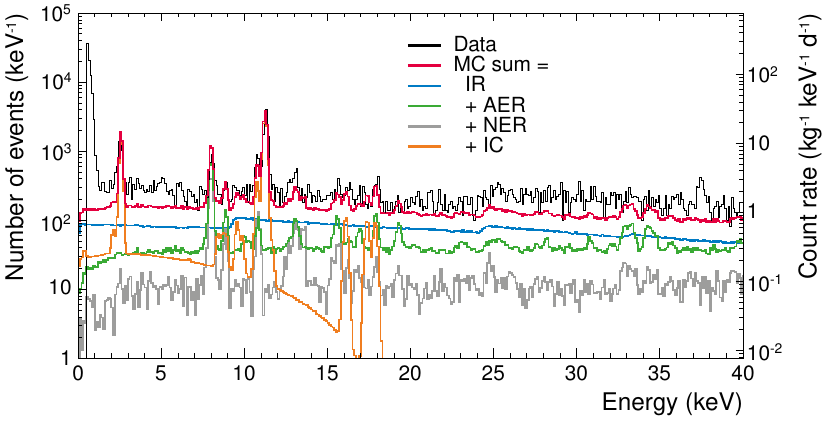}
	\label{fig:breakdown_40}
}
\caption{Total simulated background spectrum (MC sum, \textit{red}) along with the individual contributions from different background components (internal radiogenic (IR, \textit{blue} histogram), additional external radiogenic (AER, \textit{green} histogram), near external radiogenic (NER, \textit{grey} histogram), internal cosmogenic (IC, \textit{orange} histogram) below $74\,\mathrm{keV}$) with respect to the experimental data (\textit{black} histogram) in the medium-energy range \protect\subref{fig:breakdown_2800}, the low-energy range \protect\subref{fig:breakdown_500}, and the ROI $(1-40\,\mathrm{keV})$ \protect\subref{fig:breakdown_40}. The bin sizes are $1\,\mathrm{keV}$, $100\,\mathrm{eV}$, and $100\,\mathrm{eV}$, respectively.}
\label{fig:breakdown}
\end{figure*}

\begin{table}
\begin{threeparttable}
\centering
\caption{Modelled activities $A$ for individual background components over the full energy range  and the resulting background rate $R$ in the ROI of TUM40.  With linearly summed up uncertainties for the total sum. Rounded to two significant digits on the uncertainty.}
\label{tab:sum_activity}
\begin{tabular*}{\columnwidth}{@{\extracolsep{\fill}}lll@{\hskip 3pt}c@{\hskip 3pt}S[table-format=1.5(5)]}
\hline
\multicolumn{3}{l}{\multirow{2}{*}{Component}}   & $A \pm \delta_\mathrm{stat} \pm \delta_\mathrm{sys}$ & \text{$R \pm (\delta_\mathrm{stat} + \delta_\mathrm{sys})$}\\
 & &                                          & ($\mathrm{\upmu Bq \, kg^{-1}}$) & \text{($\mathrm{kg^{-1}keV^{-1}d^{-1}}$)} \\\hline
\multicolumn{3}{l}{Internal} & & \\
& \multicolumn{2}{l}{Radiogenic} & & \\
& & $^{238}\mathrm{U}$ decay chain     &  4850  $\pm$ 220 $\pm$ 130 & 0.3267 \pm 0.0016\\
& & $^{235}\mathrm{U}$ decay chain     &  1256 $\pm$ 82 $\pm$ 64 & 0.2237 \pm 0.0012\\
& & $^{232}\mathrm{Th}$ decay chain    &   155 $\pm$ 11 $\pm$ 21 & 0.01843 \pm 0.00027\\
& \multicolumn{2}{l}{Cosmogenics}      &   458 $\pm$ 2 $\pm$  40 & 0.4123 \pm 0.0044  \\\hline
\multicolumn{3}{l}{External radiogenic} & & \\
& \multicolumn{2}{l}{Near\tnote{a}}             &   900 $\pm$ 250 & 0.1362 \pm 0.0053\\ 
& \multicolumn{2}{l}{Additional}       &  3290 $\pm$ 360 $\pm$ 320 & 0.3848 \pm 0.0073\\ \hline
\multicolumn{3}{l}{Total sum} & 10910 $\pm$ 920 $\pm$ 570 & 1.502 \pm 0.015\\
\hline
\end{tabular*}
\begin{tablenotes}
     \item[a] For the activities of near external radiogenic contaminants, the values we use are upper limits. Hence, there are no associated systematic uncertainties and we give only the statistical one.
\end{tablenotes}
\end{threeparttable}
\end{table}
\begin{table}
\centering
\caption{Modelled relative contribution $\zeta$ for individual background components of TUM40 in comparison to previous work \cite[Table 3]{Strauss:2014aqw} in the energy ROI. Rounded to two significant digits on the uncertainty.}
\label{tab:comparison}
\begin{tabular*}{\columnwidth}{@{\extracolsep{\fill}} l S[table-format=2.2(2)] S[table-format=2.1(2)]}
\hline
\multirow{2}{*}{Component} & \multicolumn{2}{c}{Relative contribution $\zeta$ (\%)}\\
                            & {Previous work \cite{Strauss:2014aqw}}& {This work} \\
\hline
Internal radiogenic         & 30.4 \pm 8.1                         & 26.6 \pm  5.0\\
Internal cosmogenic         & 17.9 \pm 1.1                         & 17.8 \pm  3.8\\
Near external radiogenic    & 16.9 \pm 9.4                        &  6.3 \pm  2.1\\
Additional external radiogenic & 3.83 \pm 0.49                      & 17.5 \pm  4.9\\ \hline
Total sum                   & 69.0 \pm 9.2                                   & 68.2 \pm 15.8\\
\hline
\end{tabular*}
\end{table}
Table \ref{tab:sum_activity} lists the activities $A$ and background rates $R$ of individual background contributions as result of our Geant4 model and Fig.~\ref{fig:breakdown} shows the related background spectra. The activities represent the background that reaches the crystal as it would be seen by an ideal detector over the full energy range, i.e.\ the combined energy ranges of the low, medium, and high reference data sets. Considering the real signal survival probability $\varepsilon$ for the ROI result in the background rates $R$ which are directly comparable to the experimentally observed $R_\mathrm{TUM40}$ (see Section~\ref{reference_data}). Normalising the simulated $R$ to $R_\mathrm{TUM40}$ yield the relative background contributions $\zeta$ which are listed in Table \ref{tab:comparison} in comparison to a previous work \cite{Strauss:2014aqw}.\par
The contribution of $6.3\,\mathrm{\%}$ from the near external radiogenic background is based on the assumption that the copper contamination reaches its maximal allowed value (see Section~\ref{near_external_radiogenic}). In the opposite case, i.e.\ that the Cu contamination is negligible, our model coverage drops from maximal $68\,\mathrm{\%}$ in the ROI to roughly $62\,\mathrm{\%}$.\par
We note that the contribution of the ``near external radiogenic contamination'' and any ``additional external contamination'' changed between the previous study \cite{Strauss:2014aqw} and this work (see Table.~\ref{tab:comparison}): Whereas in \cite{Strauss:2014aqw} the near radiogenic component\footnote{Called ``external $\beta$'' in \cite[Table 3]{Strauss:2014aqw}.} was unconstrained, this work fixed it to the specific activity of the copper (see Section~\ref{near_external_radiogenic}). Consequently, the remaining observed background was shifted to the additional external component. As the contamination in the near copper parts is based on upper limits from literature, the partitioning between near and additional external radiogenic background components may change with the outcome of the ongoing copper assay or/and by including in the simulation more materials as sources of additional background.\par
It can be seen that internal background is the most prominent component in the ROI which is mostly dominated by radiogenic contaminants. These results indicate that a further purification of the $\mathrm{CaWO_{4}}$ crystal, as planned for CRESST-III Phase 2 \cite{Angloher:2015eza} is an important step towards the reduction of the overall background in the experimental data.\par
With the first completely Geant4-based electromagnetic background model for the CRESST experiment, we are able to identify and reproduce up to $(68 \pm 16)\,\mathrm{\%}$ of the background observed using the TUM40 detector module of CRESST-II. This is a methodological independent verification of the $(69.0 \pm 9.2)\,\mathrm{\%}$ obtained in the previous work \cite{Strauss:2014aqw} which is based on a combination of Geant4-based simulation and a semi-empirical fit to the background data itself. Furthermore, in this work we used all data recorded with TUM40, resulting in a four times higher statistics in the experimental reference data.\par
Beside this quantitative confirmation, this work features also qualitative improvements over \cite{Strauss:2014aqw}:
the simulation of the complete natural decay chains for the intrinsic background component allows us to quantify the background contribution for each nuclide (see Tables~\ref{tab:U238}, \ref{tab:U235}, \ref{tab:Th232}). Using Geant4 version 10.2 patch 1 allow us to simulate the the complete $\mathrm{^{234}Th}$-decay, whereas in \cite{Strauss:2014aqw} we had to rely on a simplifying workaround as we had to use Geant4 version 10.1. Additionally, we implemented the full geometry of the TUM40 detector module, whereas in \cite{Strauss:2014aqw}, only the target crystal was implemented. Hence, in this work, features like the copper fluorescence peak and the external annihilation peak occur naturally in the simulated spectra which had to be added ad hoc or were missing in \cite{Strauss:2014aqw}. These aspects highlight the advantages of using a fully Monte Carlo based model over a semi-empirical model.

\section{Summary and outlook}\label{sec:summary}
Electromagnetic background poses a significant problem for the direct DM search with the CRESST experiment at low energies where the discrimination power between $\beta$/$\gamma$-events and nuclear recoil-events decreases. To be able to observe a possible DM signal, it is imperative to understand and quantify the background. For this reason, Monte Carlo simulations using a Geant4-based toolkit were performed.\par
Based on the origin of the contaminants, we divided the background into two categories: internal (inside the $\mathrm{CaWO_{4}}$ target crystal) and external. Internal background includes $\alpha$- and $\beta$/$\gamma$-decays of nuclides from the natural decay chains of $\mathrm{{}^{232}Th}$, $\mathrm{{}^{235}U}$ and $\mathrm{{}^{238}U}$ as well as cosmogenically activated nuclides ($^{179}\mathrm{Ta}$, $^{181}\mathrm{W}$ and resulting $^{3}\mathrm{H}$). External background includes $ \beta$/$\gamma$-decays from the same natural decay chains and $^{40}\mathrm{K}$,
but originating from copper parts close to the target crystal, e.g.\ target holder, as well as distant additional parts, e.g.\ thermal copper shieldings of the cryostat. We scaled the simulated spectra of energy deposits to the activity values obtained from experimental reference data sets. This work, the first fully simulation-based background model for the CRESST experiment, shows that we can reproduce up to $(68 \pm 16)\,\mathrm{\%}$ of the observed background in the ROI.\par
The difference between the contributions from internal and external parts demonstrates that the leading contribution to the background comes from the $\mathrm{CaWO_{4}}$ target crystal itself. Based on this result, further purification of the target crystals should remain a high priority for the future. The second most prominent background originates from additional background sources around the target crystal. Hence, more precise knowledge of contamination levels of the used materials will result in a more precise background model.\par
Consequently, we prepare a dedicated measurement campaign for copper, scintillating foil and silicon-on-sapphi\-re as light detector used in the experiment. In addition we refine the study of cosmogenic activation of $\mathrm{CaWO_4}$ to identify potential further cosmogenically produced contaminants. Further improvements are expected from the simulation of the complete set of contaminants in a detailed model of the cryostat and of the shielding, in addition to the highly accurate simulation of the detector module itself which was used in this work. Beyond these studies of electromagnetic background components, we also start an in-depth investigation of the neutron background.

\section{Acknowledgements}
We are grateful to LNGS for their generous support of CRESST. This work has been supported through the FWF: I3299-N36 and W1252-N27, the DFG by the SFB1258 and the Excellence Cluster Universe, by the BMBF: 05A17WO4 and 05A17VTA and by the APVV: 15-0576.

\appendix
\section{Tables}\label{appendix}
Tables \ref{tab:alpha_activities}, \ref{tab:U238}, \ref{tab:U235}, \ref{tab:Th232}, \ref{tab:gamma_activities}, \ref{tab:cosmogenic_activities} list the individual contributions of the investigated background components.
\renewcommand{\arraystretch}{1.3}
\begin{table*}
\centering
\newcolumntype{d}[3]{D{.}{\cdot}{#1} }
    \caption{Gaussian fit values (mean $E$, amplitude $A$ and variance $\sigma$) of the alpha lines observed in the high energy reference data of TUM40 for the $\mathrm{CaWO_4}$ crystal. The number of observed events ($N_{\alpha, \mathrm{obs}}$) and the corresponding activities ($A_{\alpha}$) are calculated from the integral of the fit. The statistical uncertainties for $N_{\alpha, \mathrm{obs}}$ and $A_{\alpha}$ are calculated using the fit values as described in the text.}
    \label{tab:alpha_activities}
\begin{threeparttable}
    \begin{tabular}[c]{ | c | S[table-format=4.2(3)] | S[table-format=3.2(3)] | S[table-format=2.2(3)] | S[table-format=5.0(3)] | c | }
    \hline
    \text{Nuclide}      & {\text{E (keV)}}  & {\text{Amplitude}} & {\text{$\mathrm{\sigma}$ (keV)}}    & {\text{$N_{\mathrm{\alpha,obs}}$ ($\#$)}} & {\text{$\mathrm{A_{\alpha}}$ $\pm$ $\delta_{\text{stat}}$ $\pm$ $\delta_{\text{sys}}$ ($\mathrm{\upmu Bq \, kg^{-1}}$)}} \\ \hline
   $\mathrm{^{232}Th}$ & 4081.6  \pm 0.9   &    4.9   \pm 0.6    &   8.7  \pm 0.6         & 106   \pm 15       &   10.93 $\pm$ 0.09 $\pm$ 1.52      \\
    $\mathrm{^{238}U}$  & 4270.42 \pm 0.09  &  426.7  \pm 5.2    &   9.74 \pm 0.07        & 10420 \pm 150      & 1073.2 $\pm$ 8.7 $\pm$ 15.4     \\
    $\mathrm{^{235}U}$  & 4670.3  \pm 0.7   &   12.8  \pm 0.8    &  13.8  \pm 0.5         & 442   \pm 31       &   45.5 $\pm$ 0.6 $\pm$ 3.2      \\
    $\mathrm{^{230}Th}$ & 4767.1  \pm 1.0   &   11.3  \pm 0.7    &  17.6  \pm 1.0         & 497   \pm 43       &   51.2 $\pm$ 0.5 $\pm$ 4.4      \\
    $\mathrm{^{234}U}$  &         {-}       &          {-}       &         {-}            &          {-}           & 1085.5 $\pm$ 9.8 $\pm$ 14.9\tnote{a}     \\
    $\mathrm{^{226}Ra}$ &         {-}       &          {-}       &         {-}            &          {-}           &   66.1 $\pm$ 0.7 $\pm$ 5.0\tnote{a}      \\
    $\mathrm{^{231}Pa}$ & 5141.6  \pm 0.7   &   11.7  \pm 0.7    &  14.8  \pm 0.5         & 436   \pm 30       &   44.9 $\pm$ 0.5 $\pm$ 3.1      \\
    $\mathrm{^{210}Po}$ & 5402.7  \pm 1.1   &    6.7  \pm 0.5    &  19.0  \pm 0.9         & 321   \pm 27       &   33.1 $\pm$ 0.3 $\pm$ 2.8      \\
    $\mathrm{^{228}Th}$ & 5520.6  \pm 1.9   &    5.1  \pm 0.5    &  19.3  \pm 2.0         & 247   \pm 33       &   25.4 $\pm$ 0.3 $\pm$ 3.4      \\
    $\mathrm{^{222}Rn}$ & 5590.5  \pm 0.9   &   14.2  \pm 0.8    &  18.0  \pm 1.0         & 642   \pm 49       &   66.1 $\pm$ 0.7 $\pm$ 5.0      \\
    $\mathrm{^{224}Ra}$ & 5789.9  \pm 1.8   &    4.2  \pm 0.3    &  29.0  \pm 1.5         & 302   \pm 27       &   31.1 $\pm$ 0.4 $\pm$ 2.8      \\
    $\mathrm{^{223}Ra}$ & 5970.1  \pm 0.6   &   24.6  \pm 0.9    &  21.2  \pm 0.5         & 1309  \pm 55       &  134.8 $\pm$ 2.3 $\pm$ 5.6      \\
    $\mathrm{^{218}Po}$ &         {-}       &         {-}        &        {-}             &         {-}            &   66.1 $\pm$ 0.4 $\pm$ 5.0\tnote{a}      \\
    $\mathrm{^{227}Th}$ &         {-}       &         {-}        &        {-}             &         {-}            &  141.7 $\pm$ 2.3 $\pm$ 8.5\tnote{a}      \\
    $\mathrm{^{220}Rn}$ & 6407.7  \pm  2.4  &    2.02 \pm 0.23   &  25.3  \pm  2.0        & 128   \pm 18       &   13.2 $\pm$ 0.2 $\pm$ 1.8      \\
    $\mathrm{^{211}Bi}$ & 6751.1  \pm  0.5  &   30.05 \pm 0.97   &  19.0  \pm  0.4        & 1428  \pm 54       &  147.1 $\pm$ 1.6 $\pm$ 5.5      \\
    \hline
    \end{tabular}
\begin{tablenotes}
\item[a] The peaks $\mathrm{^{234}U}/\mathrm{^{226}Ra}$ and $\mathrm{^{218}Po}/\mathrm{^{227}Th}$ coincide with each other and form singular peaks. Activity of one of the two alpha-lines per peak is obtained via secular equilibrium. The activity of the other alpha-line is calculated by subtracting the activity of the line deduced via secular equilibrium from the activity of the whole peak.
\end{tablenotes}
\end{threeparttable}
\end{table*}

\renewcommand{\arraystretch}{1}
\renewcommand{\arraystretch}{1.3}
\begin{center}
\begin{table*}[h]
\centering
    \caption{An overview of the nuclides of the $\mathrm{^{238}U}$ decay chain. Decay modes, cumulative branching ratios B.R. with respect to the head of the chain, half-lifes and activities are included \cite{iaea}. Nuclides presented in the same cell are in secular equilibrium. Activities of each nuclide are given with systematic errors as deduced in this work for the $\mathrm{CaWO_4}$ crystal of TUM40.}
    \label{tab:U238}
    \begin{tabular}[c]{ | c | c | c | c | c |}
    \hline
    \text{Parent}    & \text{Mode}            & \text{B.R. ($\%$)} & \text{Half-life}                      & \text{$A$ $\pm \delta_\mathrm{stat} \pm \delta_\mathrm{sys}$ ($\mathrm{\upmu Bq \, kg^{-1}}$)}    \\ \hline
    $\mathrm{^{238}U}$ & $\alpha$                 & 100                  & $(4.468 \pm 0.006) \cdot 10^{9}$ y      & $1073.2 \pm 8.7 \pm 24.1$                  \\
    $\mathrm{^{234}Th}$ & $\beta ^{-}$            & 100                  & ($24.10 \pm 0.03$) d                    & $1073.2 \pm 64.3 \pm 15.4$                  \\ 
    $\mathrm{^{234}Pa}$ & $\beta ^{-}$            & 100                  & ($6.70 \pm 0.05$) h                     & $1073.2 \pm 99.9 \pm 15.4$                  \\ \hline
    $\mathrm{^{234}U}$  & $\alpha$                & 100                  & $(2.455 \pm 0.006) \cdot 10^{5}$ y      & $1085.5 \pm 9.8 \pm 24.7$                  \\ \hline
    $\mathrm{^{230}Th}$ & $\alpha$                & 100                  & $(7.54 \pm 0.03) \cdot 10^{4}$ y        & $51.2 \pm 0.5 \pm 4.8$                     \\  \hline
    $\mathrm{^{226}Ra}$ & $\alpha$                & 100                  & ($1600 \pm 7$) y                        & $66.1 \pm 0.7 \pm 5.7$                     \\ 
    $\mathrm{^{222}Rn}$ & $\alpha$                & 100                  & ($3.8235 \pm 0.0003$) d                 & $66.1 \pm 0.7 \pm 5.7$                     \\
    $\mathrm{^{218}Po}$ & $\alpha / \beta ^{-}$   &  100               & ($3.098 \pm 0.012$) m                   & $66.1 \pm 0.4 \pm 5.4$                     \\ 
    $\mathrm{^{214}Pb}$ & $\beta ^{-}$            & 99.98                  & ($27.06 \pm 0.07$) m                    & $66.1 \pm 4.4 \pm 5.0$                     \\ 
    $\mathrm{^{218}At}$ & $\alpha$                &  0.02               & ($1.5 \pm 0.3$) s                       & $0.014 \pm 0.003 \pm 0.001$                  \\ 
    $\mathrm{^{214}Bi}$ & $\alpha / \beta ^{-}$   &  99.98               & ($19.9 \pm 0.4$) m                      & $66.1 \pm 14.0 \pm 5.0$                     \\ 
    $\mathrm{^{210}Tl}$ & $\beta ^{-}$            &   0.021              & ($1.30 \pm 0.03$) m                     & $0.014 \pm 0.004 \pm 0.001$                  \\ 
    $\mathrm{^{214}Po}$ & $\alpha$                &  99.979              & $(164.3 \pm 2.0) \: \upmu$s               & $66.1 \pm 17 \pm 5.0$                     \\ \hline
    $\mathrm{^{210}Pb}$ & $\alpha / \beta ^{-}$   & 100                  & ($22.20 \pm 0.22$) y                    & $33.1 \pm 0.4 \pm 2.8$                     \\ 
    $\mathrm{^{206}Hg}$ & $\beta ^{-}$            & $1.9 \times 10^{-6}$ & ($8.32 \pm 0.07$) m                     & $(63.0 \pm 4.7 \pm 5.3) \cdot 10^{-6}$    \\ 
    $\mathrm{^{210}Bi}$ & $\alpha / \beta ^{-}$   & 100                  & ($5.012 \pm 0.005$) d                   & $33.1 \pm 2.3 \pm 2.8$                     \\ 
    $\mathrm{^{206}Tl}$ & $\beta ^{-}$            & $1.33\times10^{-4}$ & ($4.202 \pm 0.011$) m                   & $0.0044 \pm 0.0004 \pm 0.0004$                \\ 
    $\mathrm{^{210}Po}$ & $\alpha$                & 100                  & ($138.376 \pm 0.002$) d                 & $33.1 \pm 0.3 \pm 3.1$                     \\ \hline
    \end{tabular}
\end{table*}
\end{center}
\renewcommand{\arraystretch}{1}
\renewcommand{\arraystretch}{1.4}
\begin{center}
\begin{table*}[h]
\centering
    \caption{An overview of the nuclides of the $\mathrm{^{235}U}$ decay chain. Decay modes, cumulative branching ratios B.R. with respect to the head of the chain, half-lifes and activities are included \cite{iaea}. Nuclides presented in the same cell are in secular equilibrium. Activities of each nuclide are given with systematic errors as deduced in this work for the $\mathrm{CaWO_4}$ crystal of TUM40.}
    \label{tab:U235}
    \begin{tabular}[c]{ | c | c | c | c | c |}
    \hline
        Parent    & Mode            & B.R. ($\%$) & Half-life                      & $A \pm \delta_\mathrm{stat} \pm \delta_\mathrm{sys}$ ($\mathrm{\upmu Bq \, kg^{-1}}$)    \\ \hline
    $\mathrm{^{235}U}$ & $\alpha$                 & 100                       & $(7.04 \pm 0.01) \cdot 10^{8}$ y           & $45.5 \pm 0.6 \pm 3.7$                       \\
    $\mathrm{^{231}Th}$ & $\beta ^{-}$            & 100                       & ($25.52 \pm 0.01$) h                       & $45.5 \pm 1.7 \pm 3.2$                       \\ \hline
    $\mathrm{^{231}Pa}$ & $\alpha$                & 100                       & $(3.276 \pm 0.011) \cdot 10^{4}$ y         & $44.9 \pm 0.5 \pm 3.7$                       \\ \hline
    $\mathrm{^{227}Ac}$ & $\alpha / \beta ^{-}$   & 100                       & ($21.772 \pm 0.003$) y                     & $143.5  \pm 2.6  \pm 4.7$                      \\ 
    $\mathrm{^{223}Fr}$ & $\alpha / \beta ^{-}$   &   1.38                    & ($22.00 \pm 0.07$) m                       & $2.0  \pm 0.1  \pm 0.1$                        \\ 
    $\mathrm{^{227}Th}$ & $\alpha$                &  98.62                    & ($18.697  \pm 0.007$) d                     & $141.7  \pm 2.3  \pm 10.8$                      \\ \hline
    $\mathrm{^{223}Ra}$ & $\alpha$                &  99.9999                  & ($11.43  \pm 0.05$) d                       & $134.8  \pm 2.3  \pm 7.9$                      \\
    $\mathrm{^{219}At}$ & $\alpha / \beta ^{-}$   & $8.28 \times 10^{-5}$     & ($56  \pm 3$) s                             & $(12.4  \pm 1.3  \pm 0.4) \cdot 10^{-5}$      \\ 
    $\mathrm{^{215}Bi}$ & $\beta ^{-}$            & $7.75 \times 10^{-5}$     & ($7.6  \pm 0.2$) m                          & $(11.3  \pm 0.8  \pm 0.4) \cdot 10^{-5}$      \\ 
    $\mathrm{^{219}Rn}$ & $\alpha$                &  99.9999                  & ($3.96  \pm 0.01$) s                        & $134.8  \pm 34.2  \pm 5.6$                      \\ 
    $\mathrm{^{215}Po}$ & $\alpha / \beta ^{-}$   & 100                       & ($1.781  \pm 0.005$) ms                     & $134.8  \pm 13.8  \pm 5.6$                      \\
    $\mathrm{^{211}Pb}$ & $\beta ^{-}$            &  99.9998                  & ($36.1  \pm 0.2$) m                         & $134.8  \pm 10.3  \pm 5.6$                      \\ 
    $\mathrm{^{215}At}$ & $\alpha$                &  $2.3 \times 10^{-4}$     & ($0.10  \pm 0.02$) ms                       & $0.031  \pm 0.008  \pm 0.001$                    \\ \hline
    $\mathrm{^{211}Bi}$ & $\alpha / \beta ^{-}$   & 100                       & ($2.14  \pm 0.02$) m                        & $147.1  \pm 1.6  \pm 7.1$                      \\ 
    $\mathrm{^{207}Tl}$ & $\beta ^{-}$            &  99.724                   & ($4.77  \pm 0.03$) m                        & $146.7  \pm 11.4  \pm 5.5$                      \\ 
    $\mathrm{^{211}Po}$ & $\alpha$                &   0.276                   & ($0.516  \pm 0.003$) s                      & $0.41  \pm 0.10  \pm 0.02$                      \\ \hline
    \end{tabular}
\end{table*}
\end{center}
\renewcommand{\arraystretch}{1}
\renewcommand{\arraystretch}{1.4}
\begin{center}
\begin{table*}[h]
\centering
    \caption{An overview of the nuclides of the $\mathrm{^{232}Th}$ decay chain. Decay modes, cumulative branching ratios B.R. with respect to the head of the chain, half-lifes and activities are included \cite{iaea}. Nuclides presented in the same cell are in secular equilibrium. Activities of each nuclide are given with systematic errors as deduced in this work for the $\mathrm{CaWO_4}$ crystal of TUM40.}
    \label{tab:Th232}
    \begin{tabular}[c]{ | c | c | c | c | c |}
    \hline
        Parent    & Mode            & B.R. ($\%$) & Half-life                      & $A \pm \delta_\mathrm{stat} \pm \delta_\mathrm{sys}$ ($\mathrm{\upmu Bq \, kg^{-1}}$)    \\ \hline
    $\mathrm{^{232}Th}$ & $\alpha$                & 100    & $(1.40  \pm 0.01) \cdot 10^{10}$ y        & $10.9  \pm 0.1  \pm 1.5$                  \\ 
    $\mathrm{^{228}Ra}$ & $\beta ^{-}$            & 100    & ($5.75  \pm 0.03$) y                      & $10.9  \pm 0.2  \pm 1.5$                  \\ 
    $\mathrm{^{228}Ac}$ & $\beta ^{-}$            & 100    & ($6.15  \pm 0.02$) h                      & $10.9  \pm 1.1  \pm 1.5$                  \\ \hline
    $\mathrm{^{228}Th}$ & $\alpha$                & 100    & ($1.9116  \pm 0.0016$) y                  & $25.4  \pm 0.3  \pm 3.7$                  \\ \hline
    $\mathrm{^{224}Ra}$ & $\alpha$                & 100    & ($3.6319  \pm 0.0023$) d                  & $31.1  \pm 0.4  \pm 3.2$                  \\ \hline
    $\mathrm{^{220}Rn}$ & $\alpha$                & 100    & ($55.6  \pm 0.1$) s                       & $13.2  \pm 0.2  \pm 2.0$                  \\ 
    $\mathrm{^{216}Po}$ & $\alpha$                & 100    & ($0.145  \pm 0.002$) s                    & $13.2  \pm 3.1  \pm 1.8$                  \\ 
    $\mathrm{^{212}Pb}$ & $\beta ^{-}$            & 100    & ($10.64  \pm 0.01$) h                     & $13.2  \pm 0.6  \pm 1.8$                  \\ 
    $\mathrm{^{212}Bi}$ & $\alpha / \beta ^{-}$   & 100    & ($60.55  \pm 0.06$) m                     & $13.2  \pm 2.6  \pm 1.8$                  \\ 
    $\mathrm{^{208}Tl}$ & $\beta ^{-}$            &  35.94 & ($3.053  \pm 0.004$) m                    & $4.7  \pm 1.2  \pm 0.7$                   \\ 
    $\mathrm{^{212}Po}$ & $\alpha$                &  64.06 & $(0.299  \pm 0.002) \: \upmu$s              & $8.4  \pm 1  \pm 1.2$                   \\ \hline
    \end{tabular}
\end{table*}
\end{center}
\renewcommand{\arraystretch}{1}
\renewcommand{\arraystretch}{1.4}
\begin{table*}[h]
\centering
\newcolumntype{d}[3]{D{.}{\cdot}{#1} }
    \caption{Gaussian fit values (mean $E$, amplitude $A$ and variance $\sigma$) of gamma lines observed in the experimental data of TUM40 and attributed to the additional external radiogenic background. The number of observed events ($N_{\mathrm{\gamma,obs}}$) and the corresponding activities ($A_{\mathrm{\gamma}}$) are calculated from the integral of the fit. The statistical uncertainties for $N_{\mathrm{\gamma,obs}}$ and $A_{\mathrm{\gamma}}$ are calculated using the fit values as described in the text. Based on simulations, $\eta$ is the ratio between the activity of the photo peak $A_{\mathrm{\gamma}}$ and the complete spectrum $\mathrm{B_{total}}$.}
    \label{tab:gamma_activities}
    \begin{tabular}[c]{ | c | S[table-format=4.2(3)] | S[table-format=3.1(2)] | S[table-format=1.2(3)] | S[table-format=4.1(3)] | c | S[table-format=1.3(0)] | c |}
    \hline
   \text{Nuclide}     & \text{$E$}     & \text{Amplitude}     &  \text{$\sigma$} & \text{$N_\mathrm{\gamma,obs}$} & \text{$A_\mathrm{\gamma} \pm \delta_\mathrm{stat} \pm \delta_\mathrm{sys}$}  & \text{$\mathrm{\eta}$} & \text{$A_\mathrm{total} \pm \delta_\mathrm{stat} \pm \delta_\mathrm{sys}$}    \\
    & \text{(keV)} & & \text{(keV)} & \text{(\#)} & \text{($\mathrm{\upmu Bq \, kg^{-1}}$)} & & \text{($\mathrm{\upmu Bq \, kg^{-1}}$)}\\\hline
     $\mathrm{^{208}Tl}$  & 2615.0  \pm 0.3    &  45.6   \pm 3.2   &  5.8   \pm 0.3   &  332  \pm 56    &  30.1 $\pm$ 2.6  $\pm$ 2.5   & 0.052  & 579.4 $\pm$ 49.7  $\pm$ 48.7  \\
     $\mathrm{^{40}K}$    & 1460.6  \pm 0.3    &  53.1   \pm 5.3   &  3.4   \pm 0.3   &  228  \pm 61    &  20.7  $\pm$ 4.6 $\pm$ 2.8   & 0.11   & 191.1 $\pm$ 42.2  $\pm$ 25.7  \\
     $\mathrm{^{228}Ac}$  & 967.8   \pm 0.2    &  107.1  \pm 7.9   &  2.6   \pm 0.2   &  352  \pm 66    &  32.0 $\pm$ 2.5 $\pm$ 3.0   & 0.036  & 881.2 $\pm$ 68.0 $\pm$ 82.6  \\
     $\mathrm{^{212}Bi}$  & 726.6   \pm 0.2    &  79.4   \pm 9.1   &  1.7   \pm 0.2   &  168  \pm 50    &  15.2 $\pm$ 3.0  $\pm$ 2.3   & 0.1    & 152.6  $\pm$ 30.4 $\pm$ 22.5  \\  
     $\mathrm{^{214}Bi}$  & 608.9   \pm 0.1    &  83.6   \pm 6.5   &  1.7   \pm 0.1   &  356  \pm 37    &  32.3 $\pm$ 3.3  $\pm$ 3.4   & 0.14   & 225.8 $\pm$ 23.2  $\pm$ 23.7  \\     
     $\mathrm{^{212}Pb}$  & 239.8  \pm 0.05    &  614    \pm 26    &  1.30  \pm 0.05  &  1340 \pm 120   &  140.4 $\pm$ 6.1 $\pm$ 8.2    & 0.43   & 325.2 $\pm$ 14.1  $\pm$ 19.1  \\
     $\mathrm{^{226}Ra}$  & 185.6  \pm 0.04    &  36.3   \pm 2.3   &  0.63  \pm 0.04  &  574.0 \pm 4.8  &  60.2 $\pm$ 8.7 $\pm$ 5.0   & 0.58   & 103.6 $\pm$ 15.0  $\pm$ 8.7   \\
     $\mathrm{^{234}Th}$  & 92.2   \pm 0.04    &  33.8   \pm 2.7   &  0.45  \pm 0.03  &  377.7 \pm 4.0  &  19.9  $\pm$ 2.9 $\pm$ 2.1   & 0.027  & 744.8 $\pm$ 109.6 $\pm$ 79.7 \\
     $\mathrm{^{210}Pb}$  & 46.3   \pm 0.01    &  105.2  \pm 6.0   &  0.23  \pm 0.01  &  599.3 \pm 4.5  &  66.8 $\pm$ 5.7  $\pm$ 5.0   & 0.784  & 85.2 $\pm$ 7.3 $\pm$ 6.4   \\
    \hline
    \end{tabular}
\end{table*}
\renewcommand{\arraystretch}{1}
\renewcommand{\arraystretch}{1.4}
\begin{center}
\begin{table*}[h]
\centering
\newcolumntype{d}[3]{D{.}{\cdot}{#1} }
    \caption{Gaussian fit values (mean $E$, amplitude $A$ and variance $\sigma$) of cosmogenic X-ray lines observed in the experimental data of TUM40 for the $\mathrm{CaWO_4}$ crystal. The number of observed events ($N_{\text{X-ray,obs}}$) and the corresponding activities ($A_{\text{X-ray}}$) are calculated from the integral of the fit. The statistical uncertainties for $N_{\text{X-ray,obs}}$ and $A_{\text{X-ray}}$ are calculated using the fit values as described in the text. Based on simulations, $\eta$ is the ratio between the activities of the photo peak $A_{\text{X-ray}}$ and the complete spectrum $\mathrm{A_{total}}$.}
    \label{tab:cosmogenic_activities}
\begin{threeparttable}
    \begin{tabular}[c]{ | c | S[table-format=2.3(4)] | S[table-format=3.1(2)] | S[table-format=1.3(4)] | S[table-format=4.0(2)] | c | S[table-format=3.4(0)] | c |}
    \hline
  \text{Nuclide}          & {\text{$E$}}    & {\text{Amplitude}}   &  {\text{$\sigma$}} & {\text{$N_{\text{X-ray,obs}}$ ($\#$)}} & {\text{$A_\text{X-ray}$ $\pm$ $\delta_{\text{stat}}$ $\pm$ $\delta_{\text{sys}}$ }} & {\text{$\mathrm{\eta}$}} & {\text{$A_\mathrm{total}$ $\pm$ $\delta_{\text{stat}}$ $\pm$ $\delta_{\text{sys}}$ }}    \\ 
      & \text{(keV)} & & \text{(keV)} & \text{(\#)} & \text{($\mathrm{\upmu Bq \, kg^{-1}}$)} & & \text{($\mathrm{\upmu Bq \, kg^{-1}}$)}\\\hline
    $\mathrm{^{179}Ta (M_1)}$  & 2.620  \pm 0.007  &  163 \pm 12   &  0.094  \pm 0.006      & 384  \pm 38        &  49.8  $\pm$ 0.3 $\pm$ 5.0   &  1.            &   49.8  $\pm$ 0.3 $\pm$ 5.0     \\
    $\mathrm{^{179}Ta (L_2)}$  & 10.83  \pm 0.02   &  37  \pm 16   &  0.09   \pm 0.04       & 83   \pm 50        &  10.59  $\pm$ 0.04 $\pm$ 6.41   &  0.9997        &   10.59  $\pm$ 0.04 $\pm$ 6.41     \\
    $\mathrm{^{179}Ta (L_1)}$  & 11.333 \pm 0.004  &  408 \pm 18   &  0.097  \pm 0.006     & 991  \pm 76        &  126.2 $\pm$ 0.5 $\pm$ 9.7  &  0.9998        &   126.2 $\pm$ 0.5 $\pm$ 9.7    \\
    $\mathrm{^{179}Ta (K)}$    & 64.994 \pm 0.008  &  233.1 \pm 7.8    &  0.264  \pm 0.006      & 1541 \pm 64        &  170.6 $\pm$ 0.8 $\pm$ 7.0   &  0.9719        &   175.5 $\pm$ 0.8 $\pm$ 7.2     \\
    $\mathrm{^{181}W}$         & 73.49  \pm 0.05   &  19.6  \pm 1.9    &  0.51   \pm 0.04       & 249  \pm 30        &  27.0 $\pm$ 0.2 $\pm$ 3.2   &  0.4873        &   55.5 $\pm$ 0.4 $\pm$ 6.6     \\
    $\mathrm{^{3}H}$           &        {-}        &        {-}        &         {-}            &        {-}             &        {-}       &    {-}         &   40.7 $\pm$ 0.1 $\pm$ 4.8\tnote{a}    \\
    \hline
    \end{tabular}
\begin{tablenotes}
\item[a] The activity of $\mathrm{^{3}H}$ is calculated via the activity of $\mathrm{^{181}W}$. See text for details.
\end{tablenotes}
\end{threeparttable}
\end{table*}
\end{center}
\renewcommand{\arraystretch}{1}
\clearpage

\bibliographystyle{h-physrev}
\interlinepenalty=10000
\clearpage
\bibliography{our_refers}
\end{document}